\documentclass[10pt,twocolumn,british,tightenlines,eqsecnum,floats,aps,amsmath,amssymb,nofootinbib,superscriptaddress,prd,showpacs,showkeys]{revtex4}
\usepackage[latin9]{inputenc}
\usepackage{color}
\usepackage{babel}
\usepackage{amsmath}
\usepackage{amssymb}
\usepackage{graphicx}
\usepackage{esint}
\usepackage[unicode=true,pdfusetitle,
 bookmarks=true,bookmarksnumbered=false,bookmarksopen=false,
 breaklinks=false,pdfborder={0 0 1},backref=false,colorlinks=false]
 {hyperref}

\makeatletter
\@ifundefined{textcolor}{}
{%
 \definecolor{BLACK}{gray}{0}
 \definecolor{WHITE}{gray}{1}
 \definecolor{RED}{rgb}{1,0,0}
 \definecolor{GREEN}{rgb}{0,1,0}
 \definecolor{BLUE}{rgb}{0,0,1}
 \definecolor{CYAN}{cmyk}{1,0,0,0}
 \definecolor{MAGENTA}{cmyk}{0,1,0,0}
 \definecolor{YELLOW}{cmyk}{0,0,1,0}
}

\usepackage{color}\usepackage{euscript}\usepackage{epsfig}
\usepackage{amsfonts}\usepackage{fancyhdr}

\@ifundefined{textcolor}{}{%
 \definecolor{BLACK}{gray}{0}
 \definecolor{WHITE}{gray}{1}
 \definecolor{RED}{rgb}{1,0,0}
 \definecolor{GREEN}{rgb}{0,1,0}
 \definecolor{BLUE}{rgb}{0,0,1}
 \definecolor{CYAN}{cmyk}{1,0,0,0}
 \definecolor{MAGENTA}{cmyk}{0,1,0,0}
 \definecolor{YELLOW}{cmyk}{0,0,1,0}
}

\makeatother

\begin{document}

\title{Gravitational collapse of a homogeneous scalar field in deformed phase space}

\author{S. M. M. Rasouli}

\email{mrasouli@ubi.pt}
\affiliation{Departamento de F\'{i}sica, Universidade da Beira Interior, Rua Marqu\^{e}s d'Avila
e Bolama, 6200 Covilh\~{a}, Portugal}

\affiliation{Centro de Matem\'{a}tica e Aplica\c{c}\~{o}es (CMA - UBI),
Universidade da Beira Interior, Rua Marqu\^{e}s d'Avila
e Bolama, 6200 Covilh\~{a}, Portugal}

\author{A. H. Ziaie}

\email{ah\_ziaie@sbu.ac.ir}

\affiliation{Department of Physics, Shahid Beheshti University, G. C., Evin, 19839
Tehran, Iran}

\author{J. Marto}

\email{jmarto@ubi.pt}

\affiliation{Departamento de F\'{i}sica, Universidade da Beira Interior, Rua Marqu\^{e}s d'Avila
e Bolama, 6200 Covilh\~{a}, Portugal}

\affiliation{Centro de Matem\'{a}tica e Aplica\c{c}\~{o}es (CMA - UBI),
Universidade da Beira Interior, Rua Marqu\^{e}s d'Avila
e Bolama, 6200 Covilh\~{a}, Portugal}

\author{P. V. Moniz}

\email{pmoniz@ubi.pt}

\affiliation{Departamento de F\'{i}sica, Universidade da Beira Interior, Rua Marqu\^{e}s d'Avila
e Bolama, 6200 Covilh\~{a}, Portugal}

\affiliation{Centro de Matem\'{a}tica e Aplica\c{c}\~{o}es (CMA - UBI),
Universidade da Beira Interior, Rua Marqu\^{e}s d'Avila
e Bolama, 6200 Covilh\~{a}, Portugal}

\begin{abstract}
We study the gravitational collapse of a homogeneous
scalar field, minimally coupled to gravity, in the presence of a particular
type of dynamical deformation between the canonical momenta of the
scale factor and of the scalar field. In the absence of such a deformation,
a class of solutions can be found in the literature [R. Goswami and P. S. Joshi, arXiv:gr-qc/0410144], 
whereby a curvature singularity occurs at the collapse end state,
which can be either hidden behind a horizon or be visible to external
observers. However, when the phase-space is deformed, as implemented
herein this paper, we find that the singularity may be either removed
or instead, attained faster. More precisely, for negative values of
the deformation parameter, we identify the emergence of a negative
pressure term, which slows down the collapse so that the singularity
is replaced with a bounce. In this respect, the formation of a
dynamical horizon can be avoided depending on the suitable choice
of the boundary surface of the star. Whereas for positive values,
the pressure that originates from the deformation effects assists
the collapse toward the singularity formation. In this case, since
the collapse speed is unbounded, the
condition on the horizon formation is always satisfied and furthermore
the dynamical horizon develops earlier than when
the phase-space deformations are absent. These results are obtained
by means of a thoroughly numerical discussion.
\end{abstract}


\date{\today}


\maketitle

\section{Introduction}

\label{int}

\indent

One of the most important contemporary challenges in gravitation theory
and relativistic astrophysics is to fully describe the gravitational
collapse of a massive body from initially regular matter distributions
\cite{Frolov}. While Einstein's general theory of
relativity has been a highly successful theory in describing gravitation,
it is a well-established result that a gravitational collapse process,
governed by the Einstein field equations with physically reasonable
matter configurations, may induce a spacetime singularity to appear~\cite{Hawking}:
physical parameters such as the matter energy density and spacetime
curvatures will diverge. Among the variety of models that have been
investigated, the gravitational collapse of scalar fields have attracted
particular attention: massless as well as massive scalar fields have
been studied by applying analytical and numerical methods~\cite{Wym81}-\cite{GJM12}.
However, classical general relativity breaks down at the very late
stages of a collapse scenario, where densities and curvatures are
so extreme that quantum gravity effects may become more prominent,
therefore possibly resolving the classical singularity \cite{QGS,in}.

One such possible effect is noncommutativity between spacetime coordinates,
which was first proposed by Snyder \cite{SNY} in an effort to introduce
a short length cutoff (the noncommutativity parameter) in a Lorentz
covariant way. The aim was to improve the renormalizability properties
of relativistic quantum field theory (see \cite{noncom} and references
therein). The basic idea that lies behind noncommutativity is to take
into account the uncertainty in simultaneous measurements of any (canonical)
pair of phase space variables and their conjugate momenta. This idea
has been revived in recent years, due to strong motivations from string
and M-theories \cite{STMT} and more concretely has been proposed
in a new algebra regarding spacetime uncertainty relations derived
from quantum mechanics and general relativity that provides the framework
for noncommutative field theories in a Lorentz covariant way \cite{DFR}
(see also~\cite{DFR1} and references therein). One may also study
noncommutative theories in particle physics, owing to the interesting
predictions having been made in this area, such as IR/UV mixing and
nonlocality \cite{IRUV}, violation of Lorentz symmetry \cite{LV}
(see also \cite{phrev} on the phenomenological features of noncommutative
geometry), new physics at very short distance scales \cite{noncom}
and the equivalence between translations in noncommutative gauge theories
and gauge transformations \cite{TRN}. Noncommutative extensions of
quantum mechanical models such as the harmonic oscillator \cite{HO},
Hydrogen atom spectra \cite{HAS}, and gravitational radiation~\cite{GW}
have also been investigated in order to predict theoretical values
of the noncommutative parameter as a test bed for the experiments.

Since the advent of noncommutative field theory, the interest in this
area slowly but continuously made progress into the domain of gravity
theories. Recent progresses in noncommutative geometry imply that,
the noncommutative effects in general relativity may be taken into
account by keeping the standard form of the Einstein tensor on the
left-hand side of the field equations and introducing a modified energy-momentum
tensor as a source including noncommutative parameter, on the right-hand
side \cite{BHEva}. Several investigations have also been carried
out to verify the possible role of noncommutativity in cosmological
scenarios such as Newtonian cosmology \cite{NCNC}, cosmological perturbation
theory and inflationary cosmology \cite{NONCOMINF}, noncommutative
gravity~\cite{NONGRAV}, quantum cosmology \cite{NONCOMQC,BP04}
and noncommutativity based on generalized uncertainty principle \cite{GUP}
(see \cite{NCG} for reviews on different approaches to noncommutative
gravity). The concept of spacetime underlying the general theory
of relativity would not be sensible below the distances which are
comparable to the Planck ones, because the uncertainty principle governing
the quantum theory of gravity prohibits measurements in positions
to better accuracies than the Plank length. Since the transition from
classical to quantum mechanics requires the physical observables to
be noncommutative, it is expected that, in a transition from classical
to quantum gravity, the observables could also become noncommutative.
Thus, by extending general relativity toward noncommutative spacetime,
we may come closer to some aspects of quantum gravity. In particular,
replacing the usual canonical Poisson brackets between
physical variables, by others, with new terms, as suggested by string
theory, concerning fundamental interactions (see e.g.~\cite{STRNON}
and reference therein). Such deformations on the structure of the
phase-space \cite{Za05} have been employed as a means to convey noncommutativity
into the dynamics \cite{R.book04}-\cite{COR02}. In this work, our
objective is to investigate the gravitational collapse of a minimally
coupled scalar field $\phi$ in the presence of a specific phase
space deformation. In particular, this modification will concern the
dynamical sector involving the momenta of the scale factor $a$ and
of $\phi$. Our paper is organized as follows. In
Sec. ~\ref{Standard}, we briefly summarize some features regarding
the gravitational collapse of a minimally coupled homogeneous scalar
field \cite{JG04}, but within a Hamiltonian formalism. In Sec.~\ref{DPS},
we will obtain, still using the Hamiltonian formalism, the equations
of motion for such minimally coupled scalar field, but in the presence
of the particular dynamical deformation aforementioned. Subsequently,
we extract (numerically) a class of solutions that represent a gravitational
collapse. In particular, we discuss the implications regarding the
collapse outcome, for different ranges of the deformation parameter.
We will find that the deformation determines that an extra pressure
term may appear, changing the collapse dynamics and whether if a singularity
can be formed or not. We summarize our conclusions in Sec.~\ref{Concl}.
Appendices \ref{App.A} and \ref{App.B} provide complementary information
regarding the equations of motion in the deformed phase-space.

\section{Gravitational Collapse of a Homogeneous Scalar Field}

\label{Standard}

In this section, we briefly describe the gravitational
collapse of a homogeneous scalar field, minimally coupled to gravity.
The full detailed analysis is present in \cite{JG04}. In
this work we employ the Hamiltonian formalism, for the reason that
it will prove useful when we introduce, in the next section, the deformation
(noncommutativity) in the phase-space. Therefore,
we start with a Lagrangian density as
\begin{equation}
{\mathcal{L}}=\sqrt{-g}\left(\frac{{\cal {R}}}{2k^{2}}-\frac{1}{2}g^{\mu\nu}\phi_{,\mu}\phi_{,\nu}-V(\phi)\right),\label{eq1}
\end{equation}
where $k^{2}\equiv8\pi G$, ${\cal {R}}$ is the Ricci scalar, $g$
is the determinant of a metric $g_{\mu\nu}$ (where
the Greek indices run from zero to three) and $V(\phi)$ is a scalar
potential. For practical reasons we employ, for our gravitational
setting, a spherically symmetric homogeneous collapsing (interior)
region, which is given by the following line element as (cf. \cite{JG04},
\cite{JGS-LQ}-\cite{PL})
\begin{equation}
ds^{2}=h_{ab}dx^{a}dx^{b}+R^{2}(t,r)d\Omega^{2},\label{frw}
\end{equation}
where $h_{ab}={\rm diag}[-N^{2}(t),a^{2}(t)]$ is the line element
on the two-dimensional hypersurface, normal to the two dimensional
sphere characterized by the standard line element $d\Omega^{2}$.
$N(t)$ is a lapse function, $a(t)$ is the scale factor, and $R(t,r)=ra(t)$
is the physical radius of the collapsing star. Hence, the scalar field
must depend only on the comoving time, i.e $\phi=\phi(t)$. By substituting
the Ricci scalar associated to the metric (\ref{frw}) into the Lagrangian
density (\ref{eq1}), neglecting the total time derivative $k^{2}d(N^{-1}a^{2}\dot{a})/dt$,
the corresponding Hamiltonian reads
\begin{equation}
{\mathcal{H}}_{0}=-\frac{k^{2}}{12}Na^{-1}P_{a}^{2}+\frac{1}{2}Na^{-3}P_{\phi}^{2}+Na^{3}V(\phi),\label{H0}
\end{equation}
where $P_{a}$ and $P_{\phi}$ are the momentum conjugates associated
to the scale factor and scalar field, respectively. Therefore, the
Dirac Hamiltonian is given by
\begin{equation}
{\mathcal{H}}={\mathcal{H}}_{0}+\lambda P_{N},\label{Dirac.H}
\end{equation}
where we should note that, as the momentum conjugate to $N(t)$, $P_{N}$,
vanishes, we have therefore added the last term, $\lambda P_{N}$
as a constraint to the Hamiltonian~(\ref{H0}), in which $\lambda$
is a Lagrange multiplier.

Let us consider the ordinary phase-space structure described by the
usual (nonvanishing) Poisson brackets, as
\begin{equation}
\{a,P_{a}\}=\{\phi,P_{\phi}\}=\{N,P_{N}\}=1.\label{Poiss.eq1}
\end{equation}
The equations of motion with respect to the Hamiltonian (\ref{Dirac.H})
are%
\footnote{The dot represents derivative with respect to time.%
}
\begin{eqnarray}
\dot{a}\!\!\! & = & \!\!\!\{a,{\mathcal{H}}\}=-\frac{k^{2}}{6}Na^{-1}P_{a},\label{diff.eq1}\\
\dot{P}_{a}\!\!\! & = & \!\!\!\{P_{a},{\mathcal{H}}\}=-\frac{k^{2}}{12}Na^{-2}P_{a}^{2}+\frac{3}{2}Na^{-4}P_{\phi}^{2}\nonumber \\
 & - & 3Na^{2}V(\phi),\label{diff.eq2}\\
\dot{\phi}\!\!\! & = & \!\!\!\{\phi,{\mathcal{H}}\}=Na^{-3}P_{\phi},\label{diff.eq3}\\
\dot{P}_{\phi}\!\!\! & = & \!\!\!\{P_{\phi},{\mathcal{H}}\}=-Na^{3}\frac{dV_{\phi}}{d\phi},\label{diff.eq4}\\
\dot{N}\!\!\! & = & \!\!\!\{N,{\mathcal{H}}\}=\lambda,\label{diff.eq5}\\
\dot{P}_{N}\!\!\! & = & \!\!\!\{P_{N},{\mathcal{H}}\}=\frac{k^{2}}{12}a^{-1}P_{a}^{2}-\frac{1}{2}a^{-3}P_{\phi}^{2}\nonumber \\
 & - & a^{3}V(\phi).\label{diff.eq6}
\end{eqnarray}
We will work in the comoving gauge, that is, we fix $N=1$. Also,
to satisfy the constraint $P_{N}=0$ at all times, the secondary constraint
$\dot{P}_{N}=0$ should also be satisfied. Hence, it is straightforward
to show that Eqs.~(\ref{diff.eq1})-(\ref{diff.eq6}) give the dynamic
evolution for the system, as
\begin{equation}
H^{2}=\frac{k^{2}}{3}\left[\frac{1}{2}\dot{\phi}^{2}+V(\phi)\right]\equiv\frac{k^{2}}{3}\rho(t),\label{field.eq1}
\end{equation}
\begin{equation}
2\frac{\ddot{a}}{a}+H^{2}=-k^{2}\left[\frac{1}{2}\dot{\phi}^{2}-V(\phi)\right]\equiv-k^{2}p(t),\label{field.eq2}
\end{equation}
while the scalar field satisfies the Klein-Gordon equation
\begin{equation}
\ddot{\phi}+3H\dot{\phi}+\frac{dV}{d\phi}=0,\label{field.eq3}
\end{equation}
where $H=\dot{a}/a=\dot{R}/R$ is the rate of collapse. In addition,
$\rho$ and $p$ represent the energy density and pressure, respectively.
For $\dot{\phi}\neq0$, we can easily derive the Klein-Gordon equation~(\ref{field.eq3})
from the Eqs.~(\ref{field.eq1}) and (\ref{field.eq2}) or, equivalently,
from the conservation equation; thus, only two of the three Eqs.~(\ref{field.eq1})-(\ref{field.eq3})
are independent.

Let us rewrite Eqs.~(\ref{field.eq1})-(\ref{field.eq3}) in
a more convenient form, as
\begin{eqnarray}
 & H^{2} & \!\!\!\!=\frac{k^{2}}{3}\left[\frac{1}{2}a^{2}H^{2}\phi_{,a}^{2}+V(\phi)\right],\label{rewriten1}\\
\!\!\! & 3H^{2} & \!\!\!+2aHH_{,a}\!=\!-k^{2}\left[\frac{1}{2}a^{2}H^{2}\phi_{,a}^{2}-V(\phi)\right],\label{rewriten2}\\
\frac{dV}{da}\!\!\!\!\!\! & = & \!\!\!\!\!\!-4aH^{2}\phi_{,a}^{2}-a^{2}HH_{,a}\phi_{,a}^{2}-a^{2}H^{2}\phi_{,a}\phi_{,aa},\nonumber \\
\label{rewriten3}
\end{eqnarray}
where ``$,a$'' $\equiv\ d/da$. The above set of differential equations
have a general solution
\begin{eqnarray}
H(a)\!\! & = & \!\!\alpha\exp\left[-\frac{k^{2}}{2}\int a\phi_{,a}^{2}da\right],\label{Solvrew1}\\
V(a)\!\! & = & \!\!\alpha^{2}\left(\frac{3}{k^{2}}-\frac{a^{2}}{2}\phi_{,a}^{2}\right)\\
 & \times & \exp\left[-k^{2}\int a\phi_{,a}^{2}da\right],\nonumber
\end{eqnarray}
where $\alpha$ is an integration constant. In order to proceed, we
need to further specify the dependence of the scalar field upon one
of the other variables. Thus we take the following \textit{ansatz}
for the scalar field, which will induce a suitable gravitational collapse
dynamics:
\begin{equation}
\phi(a)=\sqrt{-2\beta}\ln(a),\label{Phiansatz}
\end{equation}
where $\beta<0$ is another constant. Applying (\ref{Phiansatz}),
we can then easily solve for the rate of collapse and the scalar field
potential, to get (we set $k^{2}=1$)
\begin{eqnarray}
H(a) & = & \alpha a^{\beta},\label{HPotential CC-a}\\
V(\phi) & = & \alpha^{2}\left(3+\beta\right){\rm exp}\left(-\sqrt{-2\beta}\phi\right),\label{HPotential CC}
\end{eqnarray}
where we require, additionally, that $\alpha<0$. From (\ref{HPotential CC-a}),
the scale factor reads
\begin{equation}
a(t)=\left[a_{i}^{-\beta}-\alpha\beta(t-t_{i})\right]^{-\frac{1}{\beta}},\hspace{4mm}t_{s}=t_{i}+\frac{a_{i}^{-\beta}}{{\alpha\beta}},\label{ScaleCom}
\end{equation}
where $t_{s}$ stands for the time at which the collapse ends in a
spacetime singularity. Let us be more concrete. The scalar field
is given by
\begin{equation}
\phi(t)\!=\!\mp\sqrt{\frac{-2}{\beta}}\ln\left[a_{i}^{-\beta}-\alpha\beta(t-t_{i})\right].\label{SFPOCom1}
\end{equation}
We then have the following expressions for the energy density and
Kretschmann invariant as
\begin{eqnarray}
\rho\!\!\! & = & \!\!\!3\alpha^{2}\left[a_{i}^{-\beta}-\alpha\beta(t-t_{i})\right]^{-2},\label{EKCom}\\
K\!\!\! & = & \!\!\!12\left[\left(\frac{\ddot{a}}{a}\right)^{2}+\left(\frac{\dot{a}}{a}\right)^{4}\right]\nonumber \\
 & = & \frac{24\alpha^{4}\left(1+\beta\left(1+\frac{\beta}{2}\right)\right)a_{i}^{4\beta}}{\left[1-(t-t_{i})\alpha\beta a_{i}^{\beta}\right]^{4}}.
\end{eqnarray}
The above class of collapse solutions has been found in \cite{JG04},
where it was shown that a spacetime singularity occurs, which can
be either hidden behind the event horizon (black hole) or visible
to the outside observers (naked singularity). It is
the causal structure of trapped surfaces and the apparent horizon,
which is the outermost boundary of the trapped region, that determines
the visibility or otherwise of the spacetime
singularity. If the trapped surfaces form prior to the singularity
formation, then the collapse scenario ends in a black hole and if
the trapped surfaces are delayed or failed to form until the singularity
formation, the regimes with extreme curvature and density may be seen
by the outside observers (naked singularity). Let us be clear and
to that aim we introduce the null coordinates
\begin{eqnarray}
d\xi^{+} & = & -\frac{1}{\sqrt{2}}\left[N(t)dt-a(t)dr\right],\nonumber \\
d\xi^{-} & = & -\frac{1}{\sqrt{2}}\left[N(t)dt+a(t)dr\right].
\end{eqnarray}
Metric (\ref{frw}) can be cast into double null form as
\begin{equation}
ds^{2}=-2d\xi^{+}d\xi^{-}+R(t,r)^{2}d\Omega^{2}.\label{DNULL}
\end{equation}
We assume, a spacetime which is time orientable and $\partial_{\pm}=\partial/\partial\xi^{\pm}$
are future pointing. The condition for radial null geodesics, $ds^{2}=0$,
shows that there exist two kinds of future pointing null geodesics
corresponding to $\xi^{+}={\rm constant}$ and $\xi^{-}={\rm constant}$
such that their expansion reads
\begin{equation}
\Theta_{\pm}=\frac{2}{R}\partial_{\pm}R.\label{expan}
\end{equation}
The expansion of radial null geodesics is a measure that the light
signals, being normal to the two-dimensional sphere, are diverging
($\Theta_{\pm}>0$) or converging ($\Theta_{\pm}<~0$). The spacetime
is said to be trapped, untrapped or marginally trapped if, respectively
\cite{TRHAT},\cite{M-Shay}
\begin{equation}
\Theta_{+}\Theta_{-}>0,~~~~\Theta_{+}\Theta_{-}<0,~~~~\Theta_{+}\Theta_{-}=0,\label{TRUNTRMTR}
\end{equation}
where the third class characterizes the outermost boundary of the
trapped region, the apparent horizon. Furthermore, the Misner-Sharp
energy may be defined as \cite{M-Shay}
\begin{eqnarray}
M(t,r) & = & \frac{R(t,r)}{2}\left[1-h^{ab}\partial_{a}R(t,r)\partial_{b}R(t,r)\right]\nonumber \\
 & = & \frac{R(t,r)}{2}\left[1+\frac{R^{2}(t,r)}{2}\Theta_{+}\Theta_{-}\right],\label{MITT}
\end{eqnarray}
which in our model reads $2M(t,r)=R(t,r)\dot{R}^{2}(t,r)$. Therefore,
the dynamical apparent horizon, which is a marginally trapped surface
in a spherically symmetric spacetime, is given by
\begin{equation}
\frac{2M}{R}=1.\label{DAHMS}
\end{equation}
From (\ref{MITT}) we then conclude that the spacetime region where
$2M/R>1(<1)$ is trapped (untrapped). For the solution (\ref{HPotential CC-a}),
we have
\begin{equation}
\frac{2M(t,r)}{R(t,r)}=r^{2}\alpha^{2}a^{2(1+\beta)}.\label{MSR}
\end{equation}
We now find that for $-1<\beta<0$, if the ratio $2M(t,r)/R(t,r)$
is less than one at the initial time, it would stay less than one
until the singular epoch and thus trapped surfaces fail to form
throughout the collapse process. For $\beta<-1$, trapped surfaces
do form and the singularity is necessarily covered by the spacetime
event horizon.

We subsequently show in the next section, by resorting to phase-space
deformation effects, that the corresponding gravitational collapse
procedure not only does not culminate in the formation of a spacetime
singularity but also exhibits a bouncing behavior, with which trapped
surfaces do not form.

\section{Effects of phase-space deformation on collapse dynamics and singularity
avoidance}

\label{DPS}

\indent

From arguments based on the Wigner quasidistribution function and
the Weyl correspondence between quantum-mechanical operators in Hilbert
space and ordinary $c$-number functions in phase-space~(see e.g.~\cite{Za05}
and references therein), it has been claimed that a deformation in phase
space can be applied as an alternative path to quantization. More
specifically, Moyal brackets, that are based on the Moyal product~\cite{BP04},\cite{COR02},\cite{LORS06},\cite{KJS06},
have been applied to introduce the deformation in the usual phase
space structure. In practice, for introducing such deformations, specific
Poisson brackets are employed, wherein noncommutative effects are
induced. However, for the purpose of tracing the effects of such noncommutativity
in gravity, a fundamental length is usually considered in the hope
of seeking for a fundamental theory upon which general relativity
and quantum theory can be consistently reconciled. The so-called Planck
scale is the scale at which gravitational effects become
comparable to the quantum ones \cite{DFR}. Such a regime with extreme
energy scale or equivalently with a tiny size scale occurs in the
very early universe and in the late stages of a typical gravitational
collapse of a dense star.

In this regard, much effort has been devoted to the concept of spacetime
noncommutativity and one of the main streams under investigation is
the $\kappa$-Minkowski spacetime \cite{7-32-star-product} so that
as it is shown in \cite{QGKappa}, it can appear in the framework
of quantum gravity coupled to matter fields. From a phenomenological
standpoint, $\kappa$-Minkowski spacetime provides a suitable playground
area for testing the predictions arising from deformed (doubly) special
relativity (DSR) theories~\cite{Ame01}-\cite{Kowa05}. In particular,
the DSR is related to the $\kappa$-deformation~\cite{Kowa02}. It
is believed that the noncommutativity introduced in this manner is
generally compatible with Lorentz symmetry~\cite{Kowa02,RVS07}.
The $\kappa$-Minkowski space is naturally introduced by concepts
based on the $\kappa$-Poincare algebra~\cite{Ame02}-\cite{Kowa05},
in which the ordinary brackets between coordinates are replaced by
\begin{equation}
\{x_{0},x_{i}\}=\frac{1}{\kappa}x_{i}.\label{k-minkowski}
\end{equation}
The parameter $\kappa=\epsilon/\zeta$, where $\epsilon=\pm1$~\cite{BAK01},
conveys the presence of the deformation (noncommutativity), with
dimension of mass in the units $c=\hbar=1$, such that one can interpret
$\kappa$ and $\zeta$ as dimensional parameters for fundamental energy
and length, respectively. Within cosmology, a few publications (see
e.g.~\cite{KS08,GSS11}) are present in the literature, using a few
such types of modifications in the phase-space structure, inspired
by relation~(\ref{k-minkowski}).

\begin{figure}
\centering\includegraphics[width=3in]{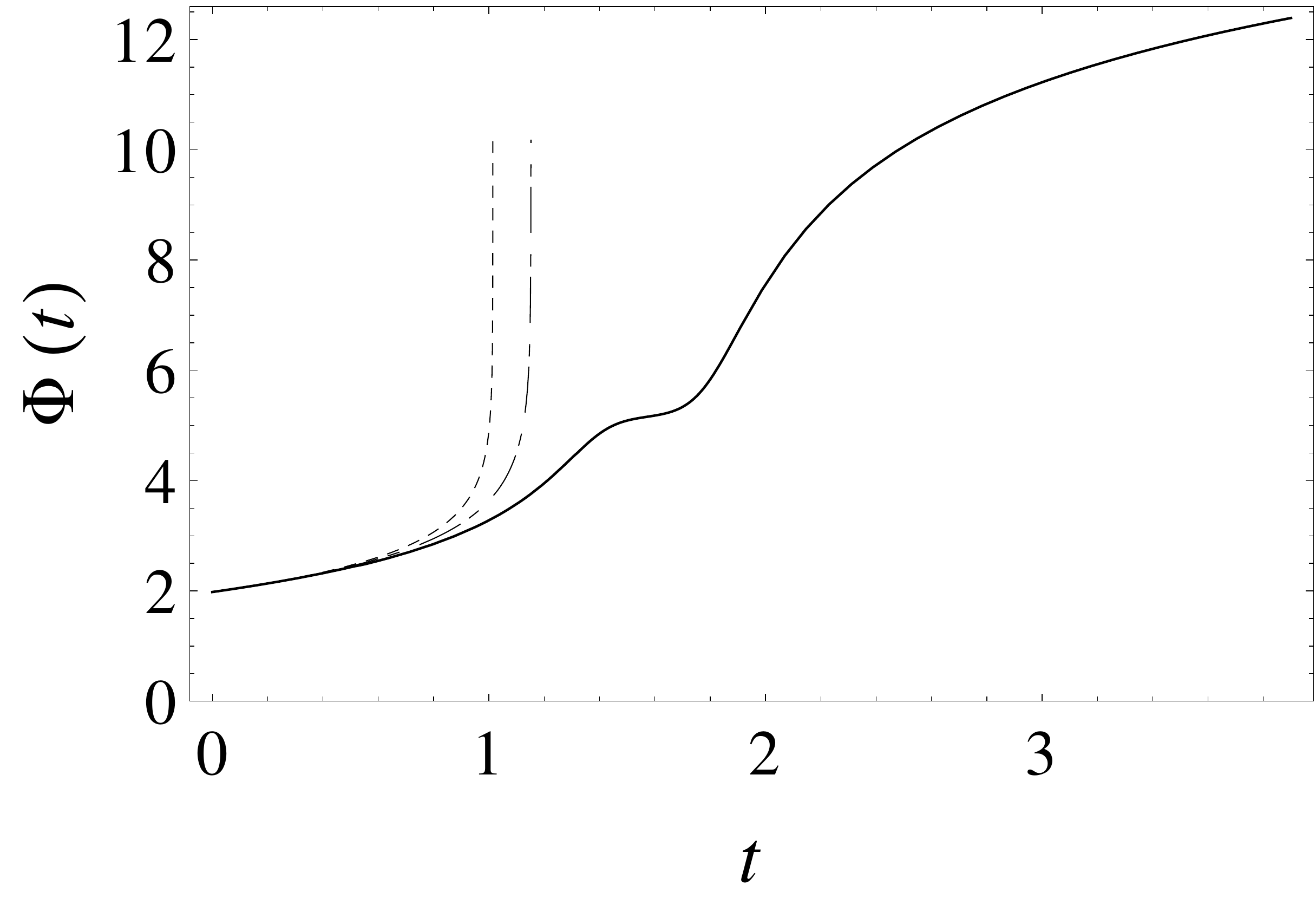}\hspace{4mm} \includegraphics[width=3in]{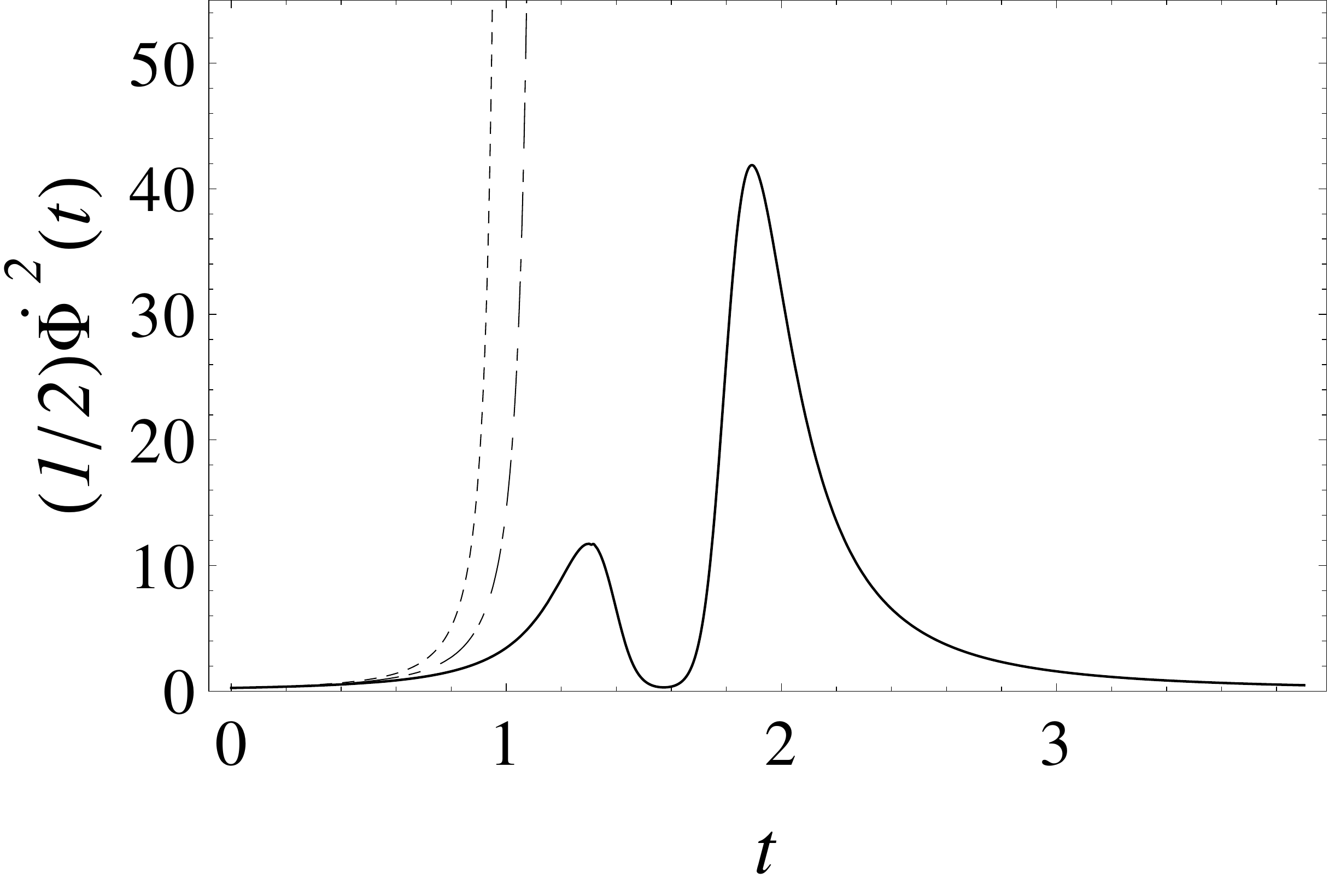}\hspace{4mm}
\caption{{\footnotesize{{The time behavior of the scalar field and its kinetic
energy for different values of deformation parameter, $\ell=-0.211$
(solid curve), $\ell=0.211$ (dashed curve), $\ell=0$ (dotted-dashed
curve) for $\beta=-3.2$ and $\alpha=1.1$. We have taken the initial
values $\phi(t_{i})=1.98$, $\dot{\phi}(t_{i})=0.711$, $a(t_{i})=3$,
$\dot{a}(t_{i})=-0.868$ and $\rho_{i}=0.2511$.}}}}

\label{DPSphi-phidot}
\end{figure}

However, in this section, inspired by the mentioned motivations in
Ref.~\cite{RFK11} and also by the corrections from string theory
to Einstein gravity~\cite{sheikh}, we propose to change the structure
of the phase-space by introducing noncommutativity between conjugate
momenta to trace the deformation implications in the gravitational
collapse of a homogeneous scalar field. To retrieve a model with deformation
(in the phase-space), where the calculations would
allow interesting novel results, but that do not convey a mere trivial
scenario, we should reasonably pick a convenient framework. Therefore,
we choose to employ a dynamical deformation within the canonical conjugate
momentum sector, {\rm viz.}, with $P_{a},P_{\phi}$ replaced by new
$P'_{a'},P'_{\phi'}$ momenta, that comply instead to
\begin{equation}
\{P'_{a'},P'_{\phi'}\}=\ell\phi'^{3},\label{deformed}
\end{equation}
where we leave the other Poisson brackets unchanged {[}corresponding
to those presented in relation (\ref{Poiss.eq1}){]}, with respect
to the above primed variables. It is straightforward to show that
the Jacobi identity is still satisfied. In~\cite{RFK11}, a discussion
on the motivations for choosing such kind of deformation in the phase
space was presented. We shall keep the Hamiltonian with the same functional
form as (\ref{H0}), but now written in terms of primed (deformed)
variables as
\begin{equation}
{\mathcal{H}'}_{0}=-\frac{1}{12}N'a'^{-1}P_{a'}'^{2}+\frac{1}{2}N'a'^{-3}P_{\phi'}'^{2}+N'a'^{3}V'(\phi\rq{}),\label{Ham.prime}
\end{equation}
where the standard Poisson brackets (for the primed variables) are
satisfied except in ~(\ref{deformed}). Here, we
aim to obtain the equations of motion for the primed variables, in
which the Dirac Hamiltonian in the deformed phase-space reads
\begin{equation}
{\mathcal{H}}^{'}={\mathcal{H}}_{0}^{'}+\lambda'P'_{N'},\label{primed-NewDirac.H-a}
\end{equation}
where ${\mathcal{H}}_{0}^{'}$ is given by (\ref{Ham.prime}) and
as $P'_{N'}=0$. We have added the last term, $\lambda'P'_{N'}$,
as a constraint to the Hamiltonian~(\ref{Ham.prime}), in which $\lambda'$
is a Lagrange multiplier and $P'_{N'}$ is the momentum conjugate
to $N'(t)$. By recalling that the deformed phase structure is described
by the deformed (nonvanishing) Poisson brackets~(\ref{deformed})
and $\{a',P'_{a'}\}=\{\phi',P'_{\phi'}\}=\{N',P'_{N'}\}=1$, the equations
of motion with respect to Hamiltonian~(\ref{primed-NewDirac.H-a})
are given by
\begin{eqnarray}
\dot{a'}\!\!\! & = & \!\!\!\{a',{\mathcal{H'}}\}=-\frac{1}{6}N'a'^{-1}P'_{a'},\label{diff.eq-prime1}\\
\dot{P'}_{a'}\!\!\! & = & \!\!\!\{P'_{a'},{\mathcal{H'}}\}=-\frac{1}{12}N'a'^{-2}P_{a'}^{'2}+\frac{3}{2}N'a'^{-4}P_{\phi'}^{'2}\nonumber \\
 & - & 3N'a^{'2}V'(\phi')+N'\ell a'^{-3}\phi'^{3}P'_{\phi'},\label{diff.eq-prime2}\\
\dot{\phi'}\!\!\! & = & \!\!\!\{\phi',{\mathcal{H'}}\}=N'a'^{-3}P'_{\phi'},\label{diff.eq-prime3}\\
\dot{P'}_{\phi'}\!\!\! & = & \!\!\!\{P'_{\phi'},{\mathcal{H'}}\}=-N'a'^{3}\frac{dV'_{\phi'}}{d\phi'}\nonumber \\
 & + & \frac{1}{6}N'\ell a'^{-1}\phi'^{3}P'_{a'},\label{diff.eq-prime4}\\
\dot{N'}\!\!\! & = & \!\!\!\{N',{\mathcal{H'}}\}=\lambda',\label{diff.eq-prime5}\\
\dot{P'}_{N'}\!\!\! & = & \!\!\!\{P'_{N'},{\mathcal{H'}}\}=\frac{1}{12}a'^{-1}P_{a'}^{'2}-\frac{1}{2}a'^{-3}P_{\phi'}^{'2}\nonumber \\
 & - & \!\!\! a'^{3}V'(\phi').\label{diff.eq-prime6}
\end{eqnarray}
Again, we work in the comoving gauge, i.e, we set $N'=1$. Also, the
constraint $P'_{N'}=0$ gives $\dot{P'}_{N'}=0$. Hence, from (\ref{diff.eq-prime6}),
we obtain
\begin{eqnarray}
P_{a'}^{'2}=6a'^{-2}P_{\phi'}^{'2}+12a'^{4}V'(\phi').\label{pa-prime}
\end{eqnarray}
By squaring both sides of Eq.~(\ref{diff.eq-prime1}) and substituting
$P_{a'}^{'2}$ from (\ref{pa-prime}) and then using Eq.~(\ref{diff.eq-prime3}),
we get the Hamiltonian constraint as
\begin{eqnarray}
\left(\frac{\dot{a'}}{a'}\right)^{2}=\frac{1}{3}\left[\frac{1}{2}\dot{\phi'}^{2}+V'(\phi')\right]\equiv\frac{1}{3}\rho'_{{\rm eff}}.\label{primed-dps-eq1}
\end{eqnarray}
Now, differentiating Eq.~(\ref{diff.eq-prime1}) with respect to
the time, and then employing Eqs.~(\ref{diff.eq-prime2}), (\ref{diff.eq-prime3}),
(\ref{pa-prime}) and (\ref{primed-dps-eq1}) we have
\begin{eqnarray}
2\frac{\ddot{a'}}{a'}+\left(\frac{\dot{a'}}{a'}\right)^{2}\!\!\! & = & \!\!\!-\left[\frac{1}{2}\dot{\phi'}^{2}-V'(\phi')\right]\\
\!\!\! & - & \!\!\!\frac{1}{3}\ell a'^{-2}\phi'^{3}\dot{\phi'}\equiv-(p'+p'_{{\rm d}})\equiv-p'_{{\rm eff}},\nonumber
\end{eqnarray}
where $p'_{{\rm d}}\equiv1/3\ell a'^{-2}\phi'^{3}\dot{\phi'}$ refers
to an effective pressure term associated to effects arising from the
deformation parameter. Finally, a modified Klein-Gordon equation can
be derived if we differentiate both sides of~(\ref{diff.eq-prime3})
with respect to time. Then, if we substitute for $\dot{P'_{\phi'}}$
from~(\ref{diff.eq-prime4}) into the resulted expression and using
relation~(\ref{diff.eq-prime1}), we extract
\begin{equation}
\ddot{\phi'}+3\left(\frac{\dot{a'}}{a'}\right)\dot{\phi'}+\frac{dV'(\phi')}{d\phi'}+\ell\dot{a'}\left(\frac{\phi'}{a'}\right)^{3}=0.\label{primed-dps-eq3}
\end{equation}
Note that, in all of the above equations, if we set $\ell=0$, then, 
each primed equation (quantity/variable) will have
the same form as its corresponding standard in the previous section.
In fact, Eqs. (\ref{primed-dps-eq1})--(\ref{primed-dps-eq3})
(where only two of them are independent), are the
extended versions of the standard equations of motion (\ref{field.eq1})--(\ref{field.eq3}).
These equations are associated to the deformed phase-space and their
solutions will describe the behavior of the primed variables. However,
hereafter, for the sake of simplicity, we drop the
prime from all the variables. We should mention that, from now on,
the herein unprimed quantities {[}which are the generalized (deformed)
forms of the standard (unprimed) ones{]} are reduced to their corresponding
in the previous section only by setting the deformation parameter
equal to zero. In Appendix~\ref{App.A}, we will present another
different approach for deriving the equations of motion. We should
note that, although not explicit, the effects of the chosen deformation
on Eq.~(\ref{primed-dps-eq3}) are implicit, as some of the
following figures will show (in particular, cf. Figs. \ref{DPS4a}
and \ref{DPSphi-phidot}, herein).

We now investigate some aspects of the gravitational
collapse, within the above framework for deformed phase-space, by
means of numerical methods. We are particularly interested in probing
the behavior of the scale factor, its time derivative, collapse acceleration,
the scalar field evolution, and other related quantities for a potential
of the same type as (\ref{HPotential CC}), in order
to properly contrast the presence of noncommutative features in
the collapse dynamics.

\begin{figure}
\centering\includegraphics[width=3.2in]{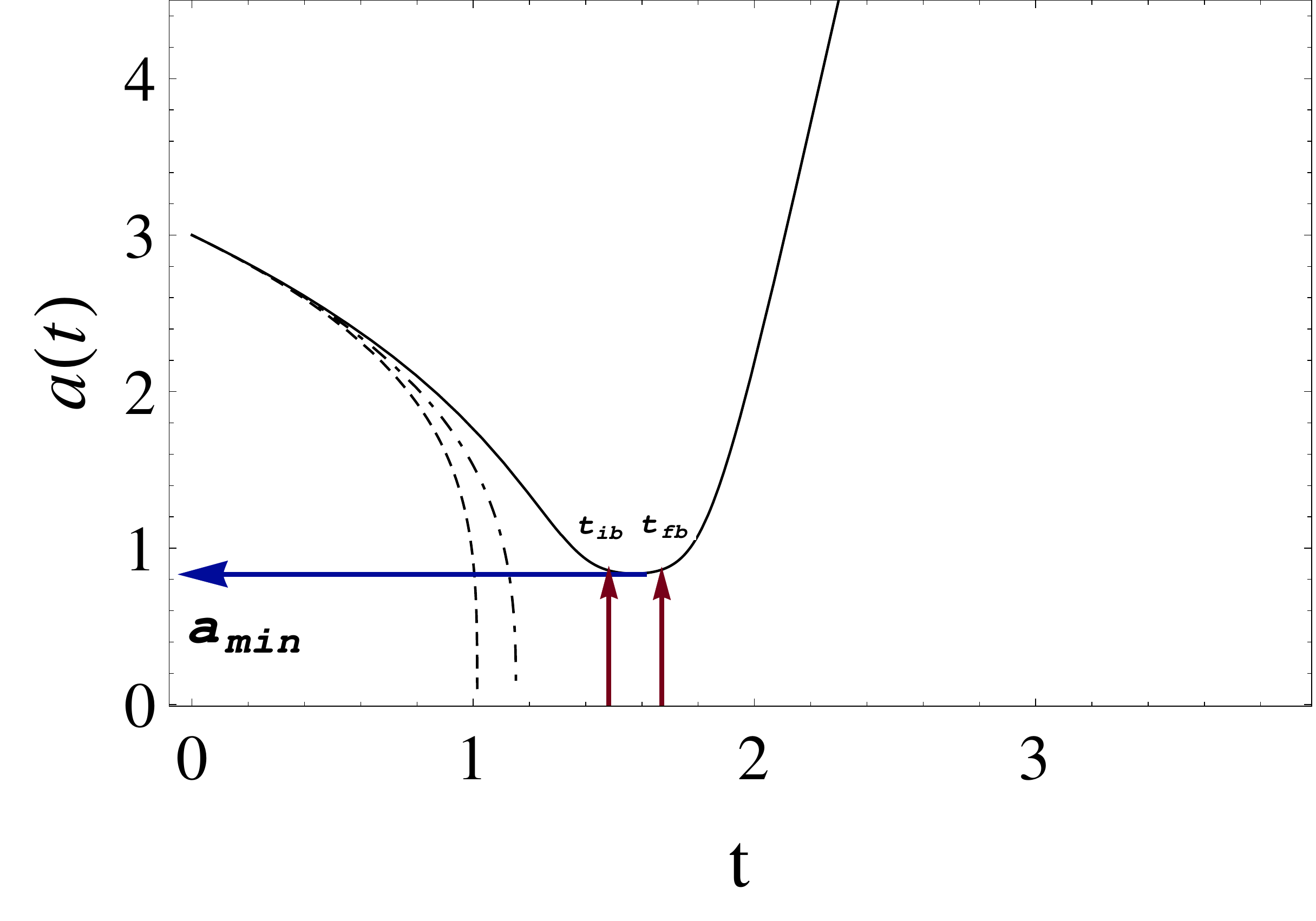} \includegraphics[width=3.3in]{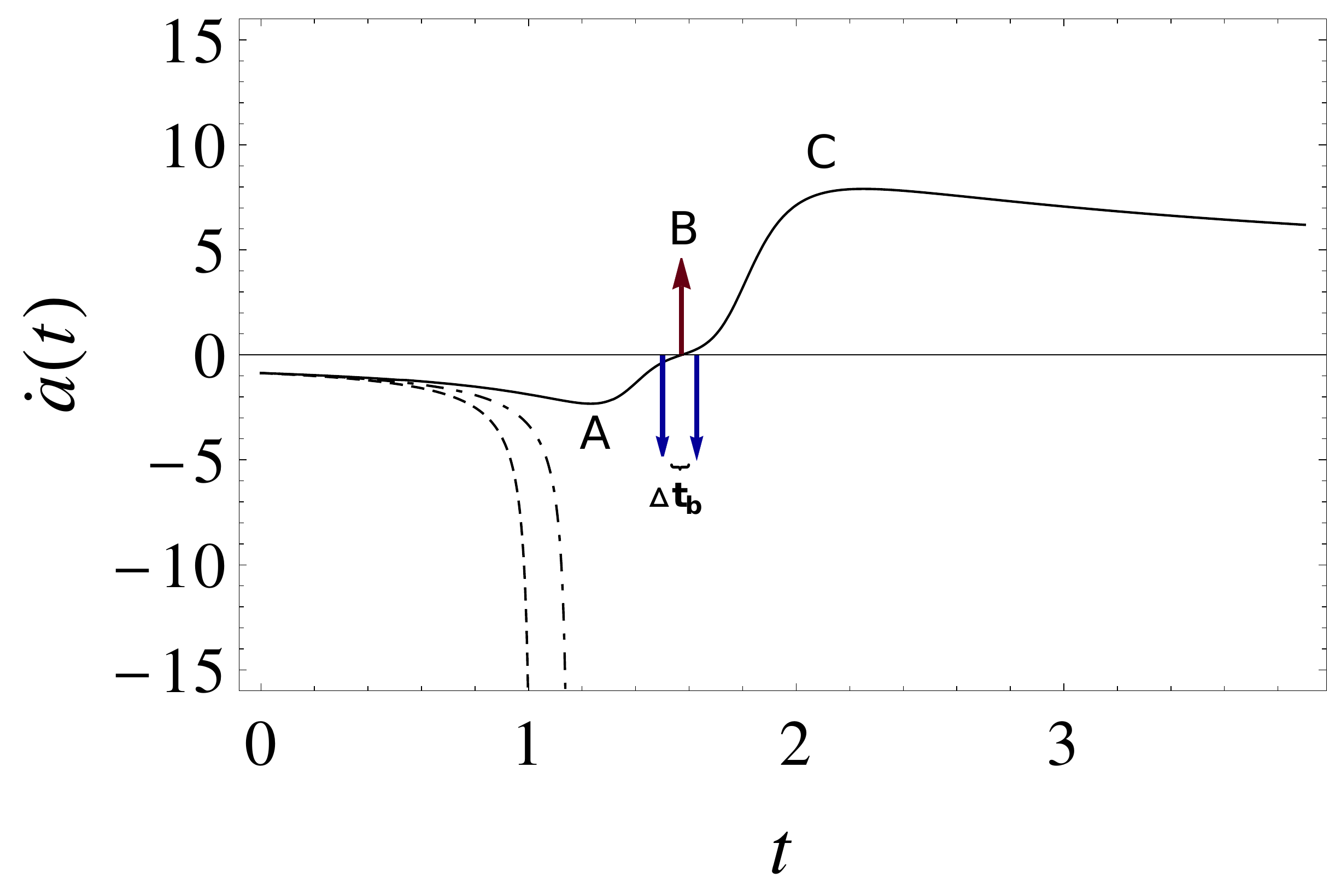}
\includegraphics[width=3.2in]{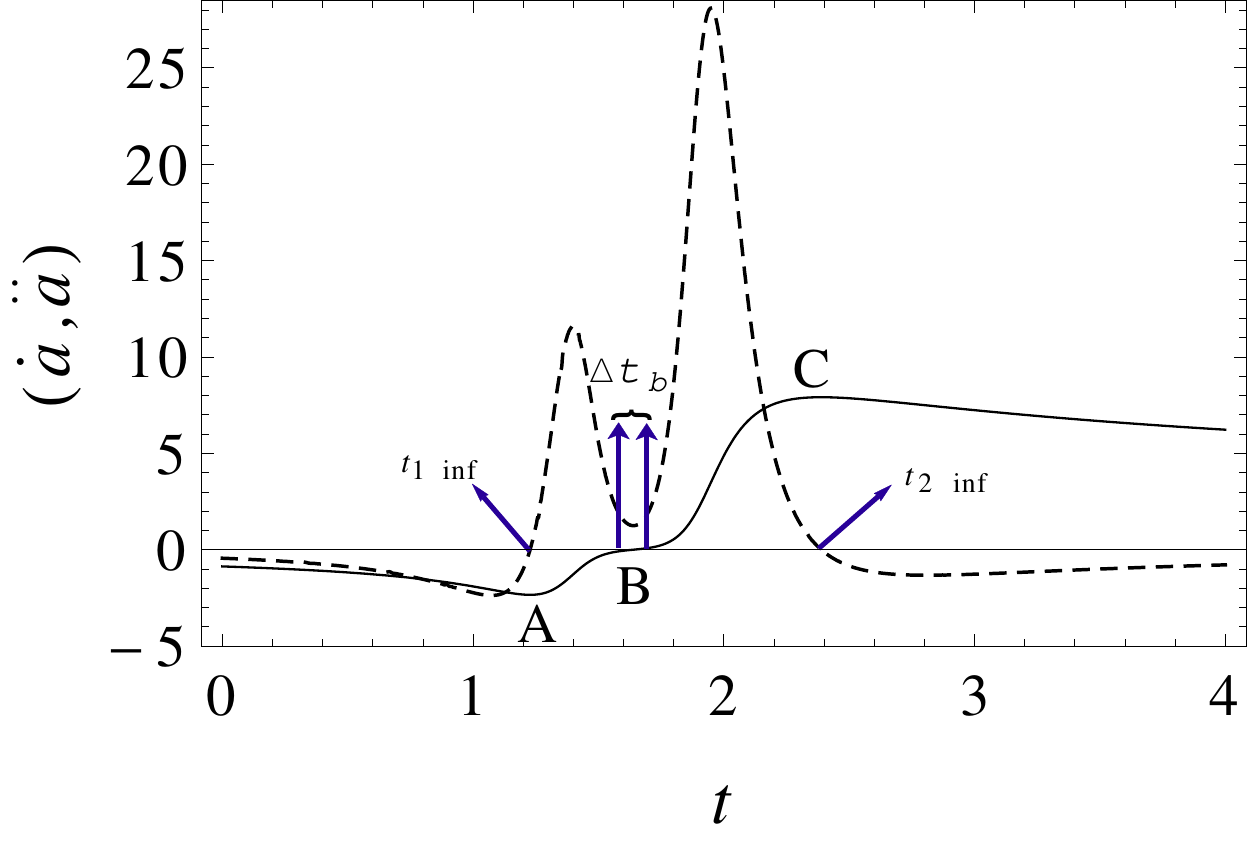} \caption{{\footnotesize{{Upper and middle panels: The time behavior of the
scale factor, the speed of collapse ($\dot{a}$) for different values
of deformation parameter, $\ell=-0.211$ (solid curve), $\ell=0.211$
(dashed curve), $\ell=0$ (dotted-dashed curve), $\beta=-3.2$ and
$\alpha=1.1$. Lower panel: The time behavior of $\dot{a}$ (solid
curve) and $\ddot{a}$ (dashed curve) for $\ell=-0.211$. We have
taken the initial values $\phi(t_{i})=1.98$, $\dot{\phi}(t_{i})=0.711$,
$a(t_{i})=3$, $\dot{a}(t_{i})=-0.868$ and $\rho_{i}=0.2511$. }}}}

\label{DPS4a}
\end{figure}

In Fig.~\ref{DPS4a}~we have presented numerically the time evolution
of the scale factor and the speed of collapse (i.e., $\dot{a}$),
for different values of the deformation parameter. All
the scale factor trajectories begin from the same initial value,
$a(t_{i})$, but, as the collapse proceeds, the full curve ($\ell<0$)
separates from the other two and reaches a minimum value for the scale
factor at a critical epoch which lies between $t_{{\rm ib}}<t_{{\rm cr}}<t_{{\rm fb}}$.
Thus, for $t_{{\rm ib}}<t<t_{{\rm cr}}$, the collapse scenario proceeds
much slower than $t<t_{{\rm ib}}$, ceasing at $t_{{\rm cr}}$
and then entering a smooth expanding phase for $t_{{\rm cr}}<t<t_{{\rm fb}}$.
Therefore, it is seen that for $\ell<0$ the collapse scenario presents
a soft bouncing behavior during the time interval $\Delta t_{{\rm b}}=t_{{\rm fb}}-t_{{\rm ib}}$.
For $\ell>0$ the collapse advances towards the singularity faster
than in the case where the phase-space deformation
effects are absent. From the middle panel of Fig.
\ref{DPS4a}, we further see that for $\ell<0$ the collapse commences
from $\dot{a}(t_{i})<0$, proceeding for a while in an accelerating
phase until an absolute maximum value in negative direction is reached
(point A). It then decelerates and halts at point B where $\dot{a}(t_{{\rm cr}})=0$.
After this epoch, the collapse regime is replaced by an accelerated
expansion and continues up to the point C. This expanding
phase slows down when this point is passed. The lower panel
in Fig. \ref{DPS4a} further supports this argument: the
collapse acceleration remains negative prior to point A, where the collapse speed
 achieves its maximum negative value. This point
corresponds to the first inflection point of acceleration curve, occurring
at $t=t_{{\rm 1inf}}$. Thus, for $t<t_{{\rm {1inf}}}$ the collapse
proceeds in the so-called \textit{fast-reacting} process while 
for $t_{{\rm {1inf}}}<t<t_{{\rm cr}}$ a \textit{slow-reacting}
regime governs. The collapse procedure experiences a decelerating
phase from points A to B (see the middle plot in Fig. \ref{DPS4a})
with $\ddot{a}$ achieving in between a local maximum. As time evolves,
the acceleration decreases to point B, with $\dot{a}$ progressing
toward less negative values (upwards), eventually being $\dot{a}\sim0$
and then smoothly becoming positive. This happens during the time
interval $\Delta t_{{\rm b}}$, within which the bounce appears. We
note that $\Delta t_{{\rm b}}$ is too small so that $\dot{a}$ changes
infinitesimally and $\ddot{a}\sim{\rm constant}$. For $t>t_{{\rm fb}}$,
an accelerating expanding phase governs the scenario until the time
$t_{{\rm 2inf}}$, at which $\ddot{a}$ reaches its second inflection
point, where $\dot{a}$ achieves its absolute maximum (see also point
C). For $t>t_{{\rm 2inf}}$ the expanding phase slows down at late
times. The situation is quite different for $\ell>0$; as it is seen
the collapse evolves faster than the case $\ell=0$.

\begin{figure}
\centering \includegraphics[width=3in]{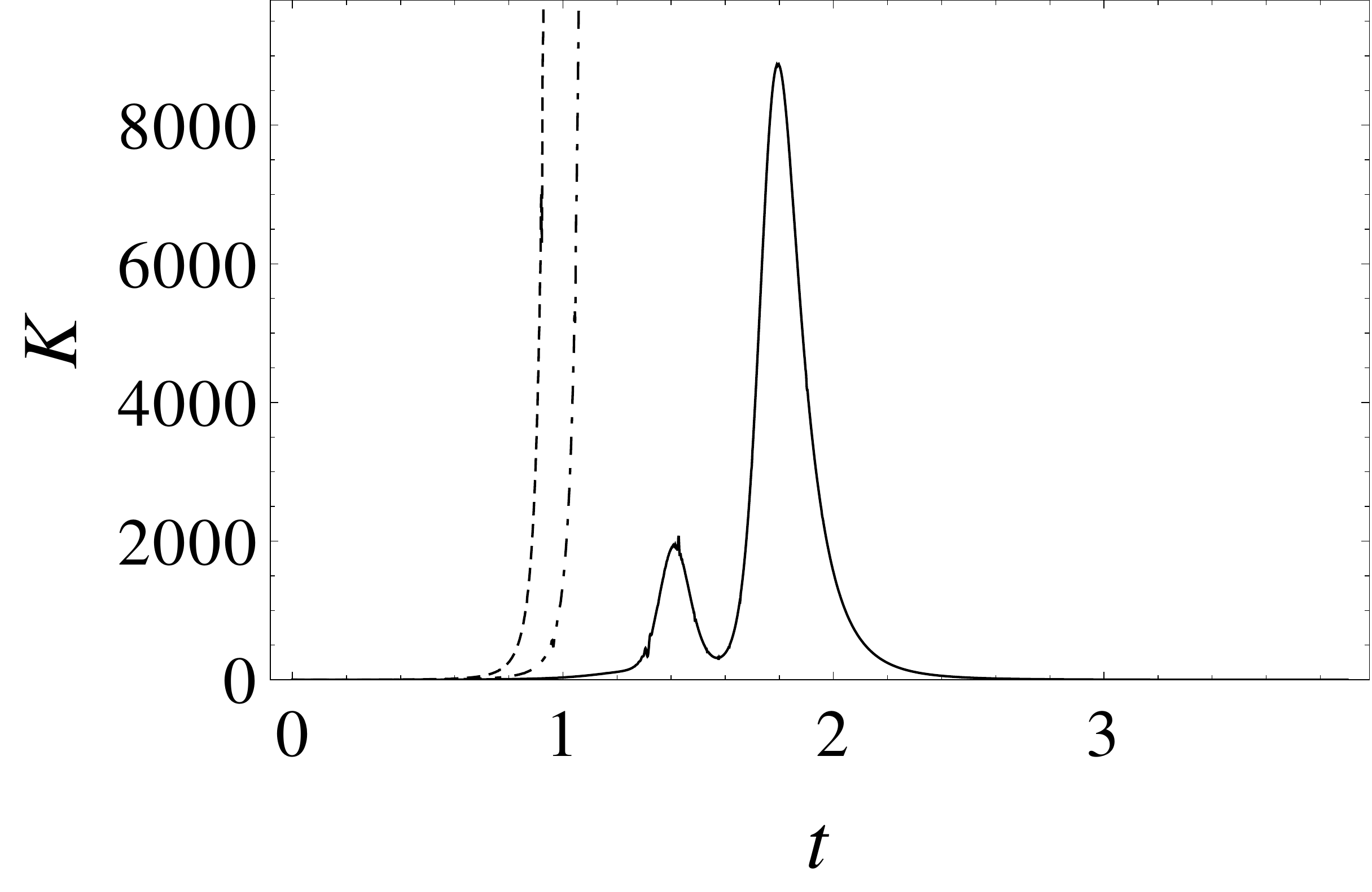} \includegraphics[width=3in]{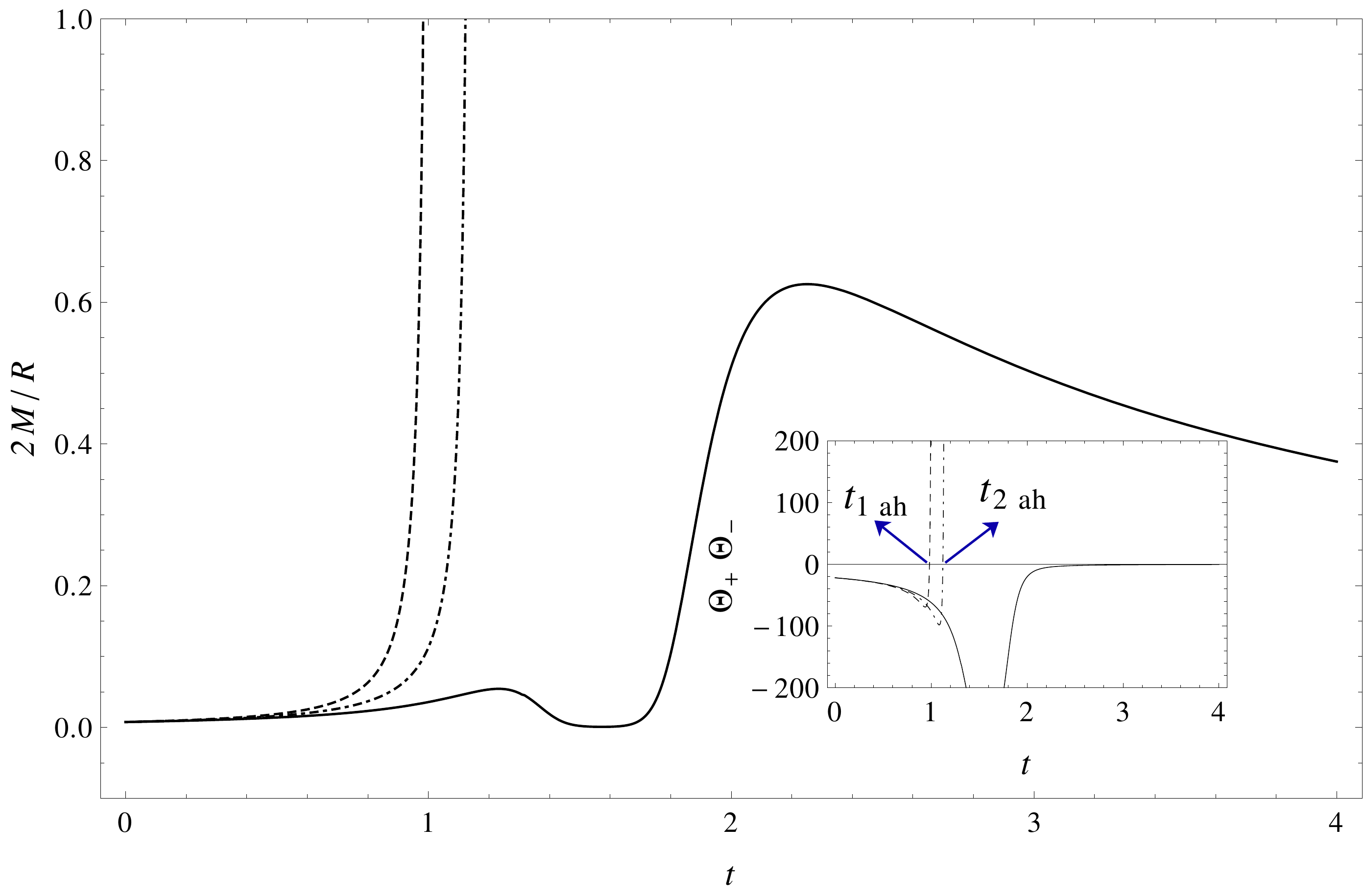}
\includegraphics[width=3in]{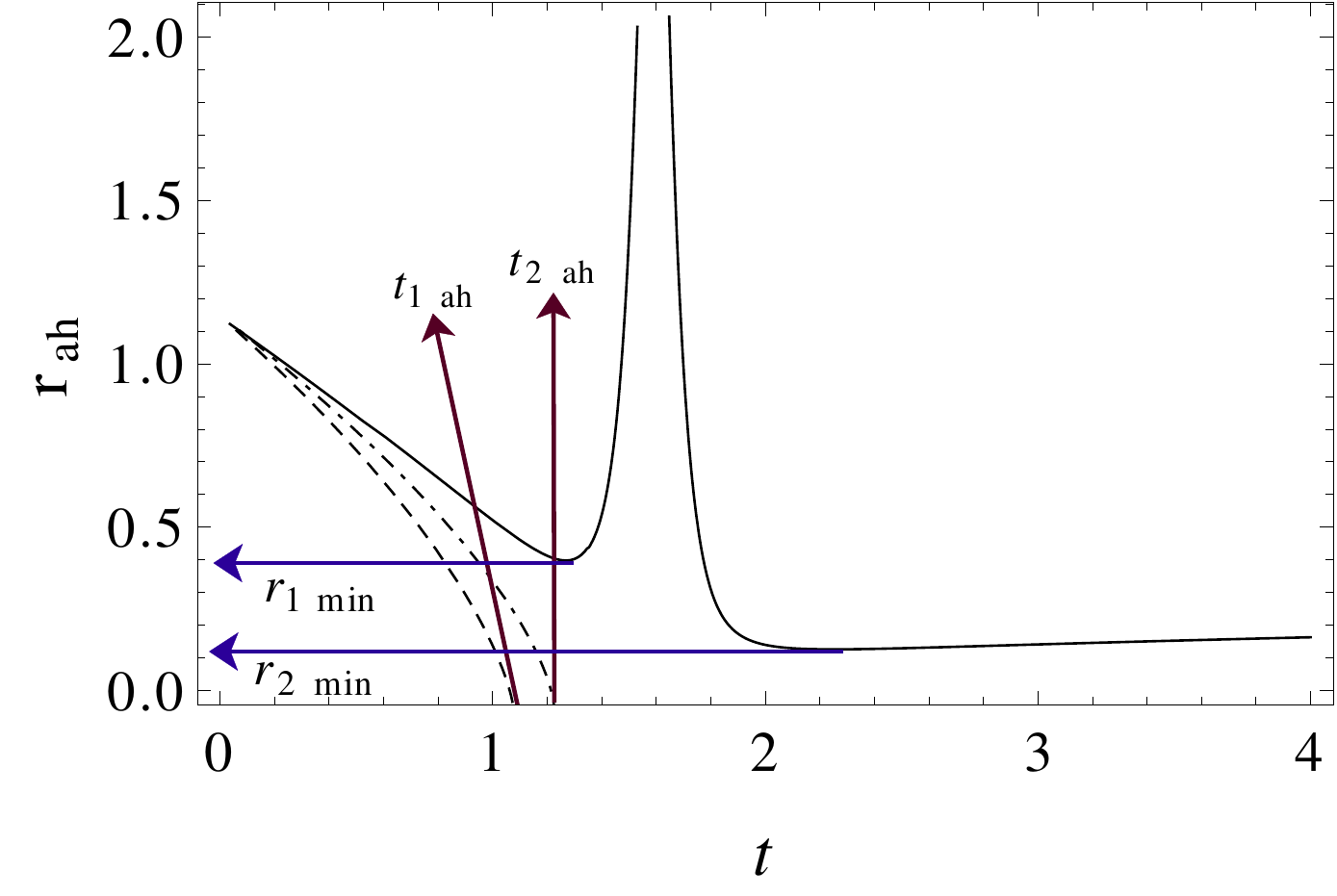} \caption{{\footnotesize{{The time behavior of the Kretschmann invariant (upper
panel), the ratio $2M/R$ (middle panel) and the apparent horizon
curve (lower panel), for different values of deformation parameter,
$\ell=-0.211$ (solid curve), $\ell=0.211$ (dashed curve), $\ell=0$ (dotted-dashed
curve), $\beta=-3.2$ and $\alpha=1.1$. The inset shows the time
behavior of the invariant $\Theta_{+}\Theta_{-}$, $\ell=-0.211$ (solid
curve), $\ell=0$ (dotted-dashed curve) and $\ell=0.211$ (dashed curve).
We have taken the initial values $\phi(t_{i})=1.98$, $\dot{\phi}(t_{i})=0.711$,
$a(t_{i})=3$, $\dot{a}(t_{i})=-0.868$ and $\rho_{i}=0.2511$.}}}}

\label{DPS4b}
\end{figure}

In Fig. \ref{DPS4b}~~we have plotted the Kretschmann invariant
(upper plot) and the ratio of twice Misner-Sharp energy (middle plot)
over the physical area radius. Correspondingly, the Kretschmann invariant
behaves regularly for $\ell<0$ but for $\ell>0$, it diverges in
a more rapid way than the case $\ell=0$. For $\ell<0$, the ratio
$2M/R$ stays finite and less than one until the bounce
occurs, which signals the trapped surfaces formation
failure; for $\ell>0$, this invariant tends to
infinity faster when compared to the case $\ell=0$:
this implies that the trapped surfaces form earlier
than when the deformation effects are absent. The
lower panel in Fig. \ref{DPS4b} further illustrates
the dynamics of the apparent horizon in the interior spacetime, which
by means of Eq. (\ref{DAHMS}), reads
\begin{equation}
r_{{\rm ah}}(t)=\frac{1}{a(t)}\sqrt{\frac{3}{\rho_{{\rm eff}}(t)}}.\label{raht}
\end{equation}
As the figure shows, the different behaviors of the scale factor bodes
the different pictures for the time behavior of the apparent
horizon curve. For $\ell<0$ (solid curve) there are two minimum radii
for which, if the boundary is taken so that $r_{{\rm 2min}}<r_{{\rm b}}<r_{{\rm 1min}}$,
the apparent horizon curve goes to infinity as $t_{{\rm cr}}$ is
approached; therefore no trapped surfaces are expected
to appear throughout the gravitational contraction
process, before the bounce occurs. However, when contraction turns
to an accelerated expansion, the apparent horizon may still form due
to the process of recapturing the mass that might have escaped during
the contraction regime. Thus, for $r_{{\rm b}}<r_{{\rm 2min}}$, no
horizon may form during the expanding phase. It is also worth noticing
that the middle plot in Fig. \ref{DPS4b} has been made for $r_{{\rm b}}=0.1$
while $r_{{\rm 1min}}\simeq0.427$ and $r_{{\rm 2min}}\simeq0.126$.
We also note that the regularity condition, which states that there
should not be any trapped surface at the initial time from which the
collapse begins, puts an upper bound on the value of the boundary.
Thus, from Eq. (\ref{raht}), the boundary has to satisfy $r_{{\rm b}}<r_{{\rm ah}}(t_{i})$
in order that the regularity condition be respected.

The case $\ell>0$ (dashed curve) also shows that there is no minimum
radius below which trapped surface formation could be avoided and
the apparent horizon forms faster than when $\ell=0$
(dotted-dashed curve). The inset of the middle panel
in Fig. \ref{DPS4b} elaborates more on this issue, where we show
the behavior of the invariant $\Theta_{+}\Theta_{-}$ over time. All
the curves begin from initial configurations that respect the regularity
condition {[}$\Theta_{+}\Theta_{-}(t_{i})<0${]}. For $\ell<0$, the
expansion of radial null geodesics stays negative throughout the scenario
which shows the failure of formation of the apparent horizon. For
$\ell=0$, this quantity stays negative for a while, then intersects
the line $\Theta_{+}\Theta_{-}=0$ at $t=t_{{\rm 2ah}}$; correspondingly,
from the lower plot of Fig. \ref{DPS4b}, we observe that the apparent
horizon (dotted-dashed curve) forms at this time to cover the singularity.
For $\ell>0$, the dashed curve gets zero at $t=t_{{\rm 1ah}}<t_{{\rm 2ah}}$,
which from the lower plot we see that the apparent horizon (dashed
curve) forms earlier in the absence of phase-space deformation effects.

\begin{figure}
\centering\includegraphics[width=3in]{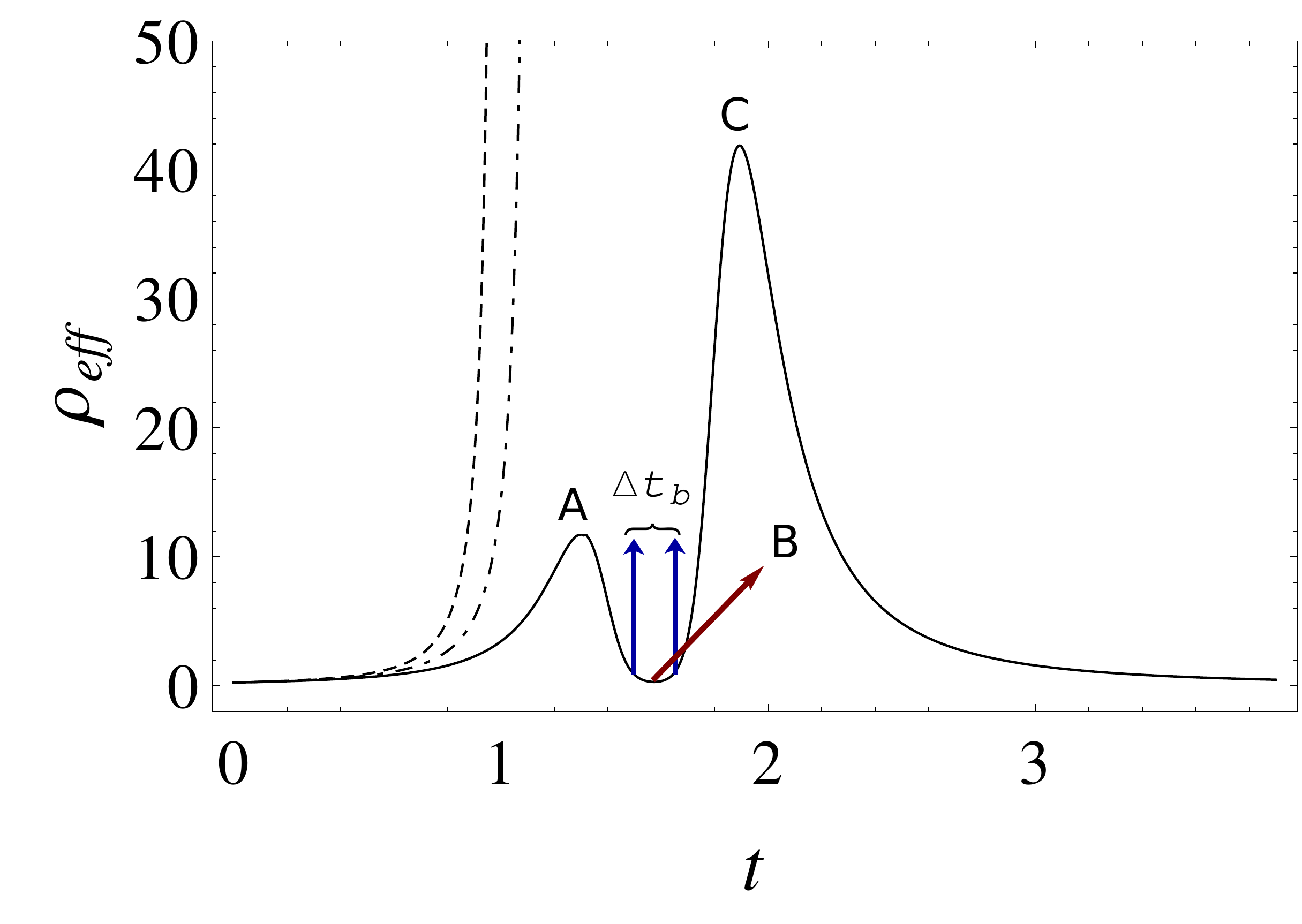}\hspace{4mm} \includegraphics[width=3in]{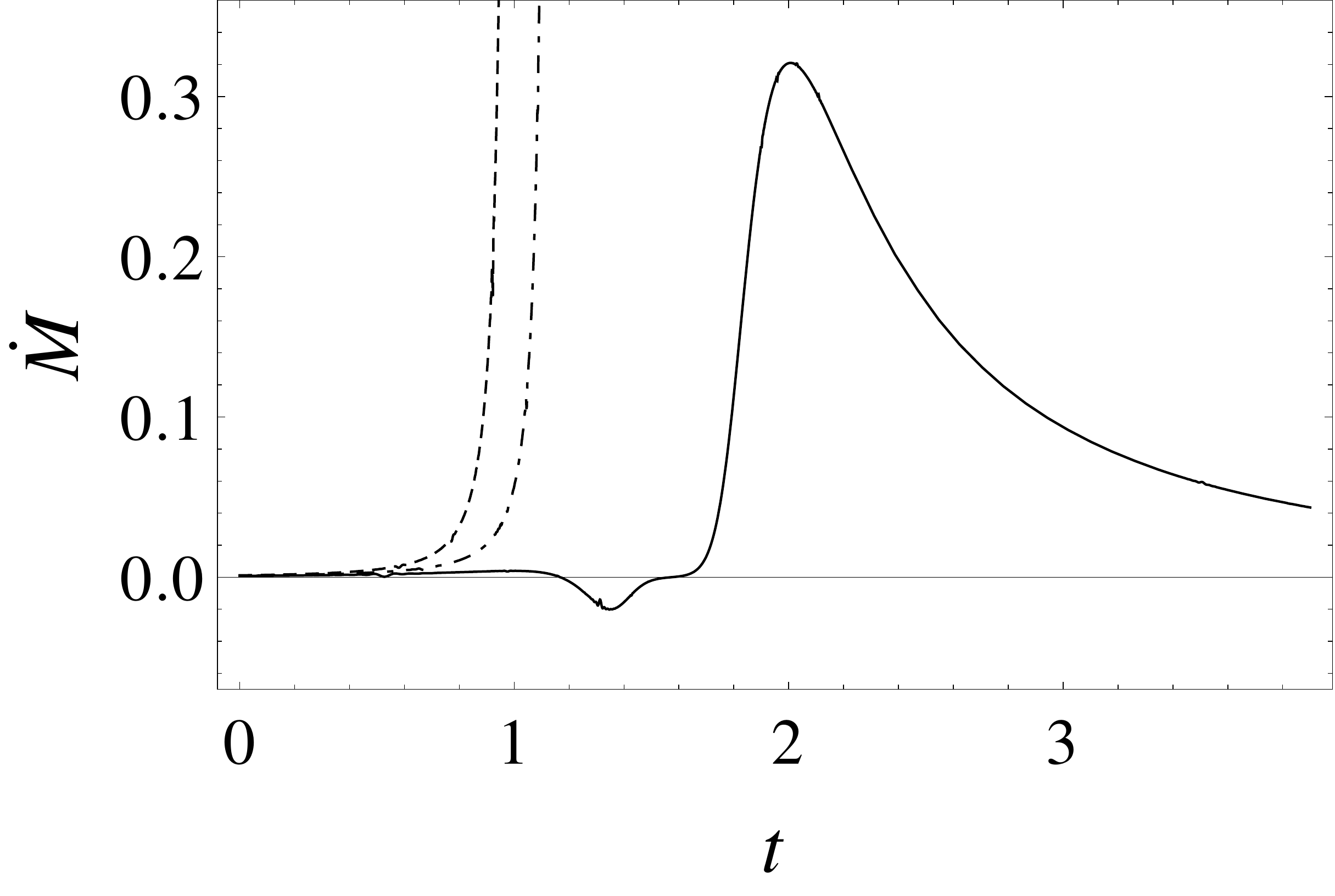}\hspace{4mm}
\caption{{\footnotesize{{The time behavior of the effective energy density
and $\dot{M}$ for different values of deformation parameter, $\ell=-0.211$
(solid curve), $\ell=0.211$ (dashed curve), $\ell=0$ (dotted-dashed
curve) for $\beta=-3.2$ and $\alpha=1.1$. We have taken the initial
values $\phi(t_{i})=1.98$, $\dot{\phi}(t_{i})=0.711$, $a(t_{i})=3$,
$\dot{a}(t_{i})=-0.868$ and $\rho_{i}=0.2511$. }}}}

\label{DPS4c}
\end{figure}

The behavior of the effective energy density and the time derivative
of the Misner-Sharp energy is shown in Fig. \ref{DPS4c} . The full
curve shows that the effective energy density increases up to a local
maximum value (point A), where $\dot{a}$ has reached its negative
maximum value. This could be associated with the first inflection
point $t_{{\rm 1inf}}$, as the star contracts in an accelerating
way. As the decelerating regime begins, the collapse slows down and
proceeds to momentarily to stop. The effective energy density decreases
to a zero value at $t=t_{{\rm cr}}$, as if some mass loss took place,
possibly caused by the appearance of negative pressure coming from
phase-space deformation effects, i.e., energy escaping as this reverse
in the dynamics takes place. Since trapped surfaces are failed to
form, this flux of energy may be visible by external
observers. Thus, a soft bounce occurs when the contracting phase transits
to an expanding one at a specific value of the scale factor between
the time interval $\Delta t_{{\rm b}}$ (point B in upper plot of
Fig. \ref{DPS4c}; cf. Fig. \ref{DPS4a}). Subsequently, as the
collapse turns into expansion, the effective energy
density increases again to an absolute maximum (point C), possibly
due to energy being regained; finally, it decreases
asymptotically, as the star continues to expand without restriction%
\footnote{A similar behavior for the energy energy density is found in the models
driven by spinor cosmology~\cite{Kibble}.%
}. The energy density never blows up and the singularity that was produced
in the undeformed case is avoided. This behavior may be interpreted
as follows: the collapse proceeds for a while and then halts at the
bouncing stage (cf. the behaviors of $\dot{a}$ and $\ddot{a}$),
after which the star expands and intakes some mass that may have escaped.
Finally, as it carries on expanding without regaining any more mass,
the density decreases. We notice that the mentioned behavior is only
for the solid line.

\begin{figure}
\centering\includegraphics[width=3in]{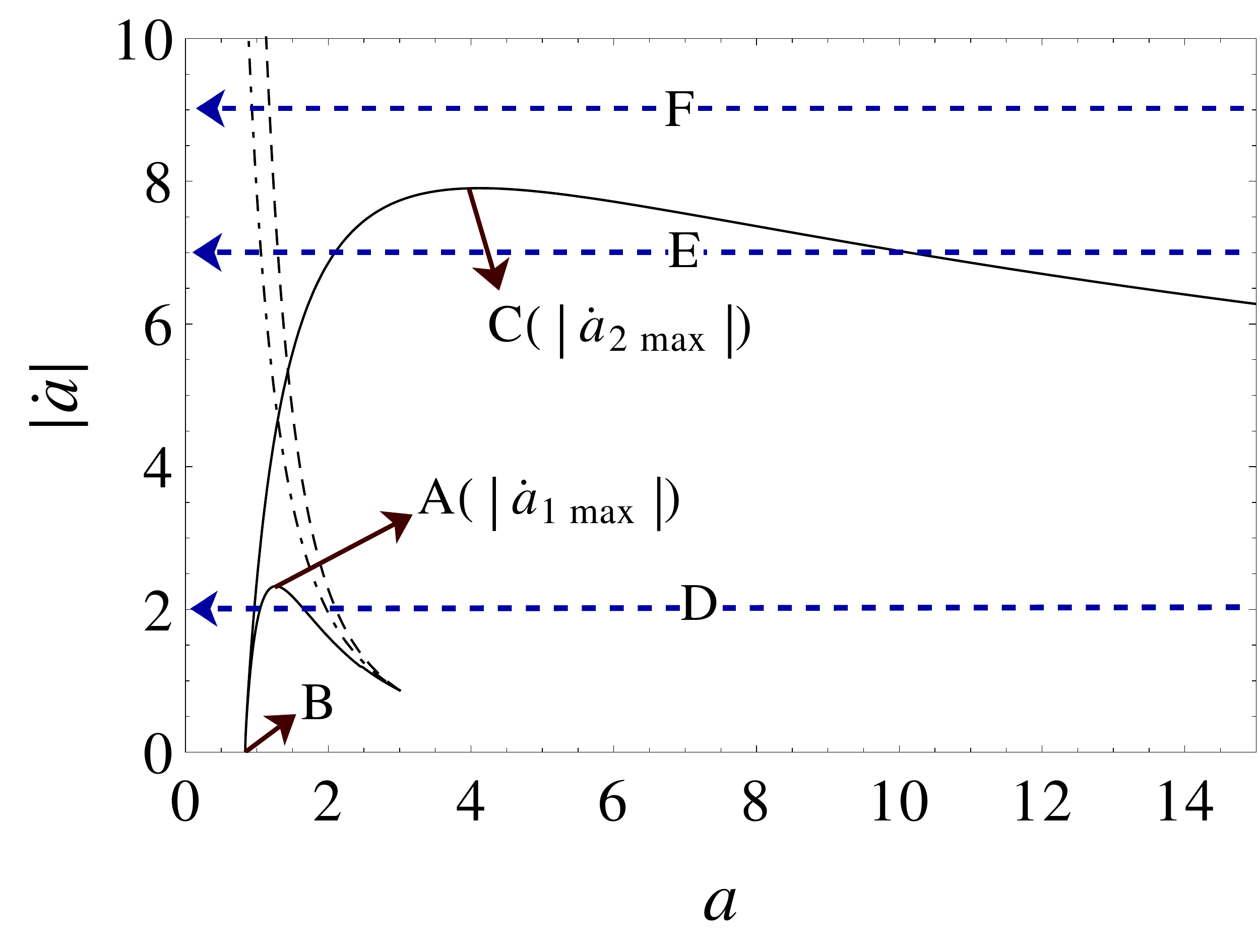} \caption{{\footnotesize{{The behavior of absolute value of the collapse velocity
in terms of the scale factor for different values of deformation parameter,
$\ell=-0.211$ (solid curve), $\ell=0.211$ (dashed curve) and $\ell=0$
(dotted-dashed curve) for $\beta=-3.2$ and $\alpha=1.1$. The horizontal
dashed arrows correspond to different values of $r_{{\rm b}}$, see
the text for details. We have taken the initial values $\phi(t_{i})=1.98$,
$\dot{\phi}(t_{i})=0.711$, $a(t_{i})=3$, $\dot{a}(t_{i})=-0.868$
and $\rho_{i}=0.2511$.}}}}

\label{DPS4ho}
\end{figure}

The lower panel in Fig.~\ref{DPS4c} suggests
indeed this behavior at this period; hence the peak for point C. Furthermore,
the negative zone in this figure shows the outward flux of energy
occurring in the deceleration phase, i.e., from points A to B in Fig.
\ref{DPS4a}. Therefore, the matching with a suitable exterior geometry,
namely the generalized Vaidya spacetime, must be carried out, which
describes an outgoing radiation. We note that since the star has internal
pressure and is radiating, the Schwarzchild metric may no longer be
a suitable spacetime to describe the exterior region. On the other
hand, through the matching procedure, the interior solution can be
extended to the exterior region%
\footnote{Correspondingly, the effects of phase-space deformation may be transported
to the outside.%
} and this may tell us whether the horizons form. Let us be more precise.
The geometry outside a spherically symmetric radiating body is given
by the generalized Vaidya metric as \cite{VaidyaM}
\begin{equation}
ds_{{\rm out}}^{2}=-\left(1-\frac{2{\mathcal{M}}(u)}{r_{v}}\right)du^{2}-2dudr_{v}+r_{v}^{2}d\Omega^{2},\label{Vaidya}
\end{equation}
where $\textit{u}=t-r_{v}$ and ${\mathcal{M}}(u)$ being the retarded
(exploding) null coordinate and the gravitational mass inside the
sphere of radius $r_{v}$, respectively. The above metric is to be
matched, by means of Isreal-Darmois junction conditions \cite{IS-DAR},
to the internal geometry [cf. (\ref{frw})] at the boundary of the star
which is a timelike hypersurface given by $r=r_{\Sigma}$. We assume
that the second fundamental form is continuous across the boundary;
there is no surface stress-energy or surface tension at the boundary
(see \cite{SETEN} for more details). The induced metrics as we approach
$\Sigma$ from the interior and exterior regions are given by, respectively
\begin{eqnarray}
ds_{\Sigma_{{\rm in}}}^{2} & = & -dt^{2}+a^{2}(t)r_{\Sigma}^{2}d\Omega^{2},\nonumber \\
ds_{\Sigma_{{\rm out}}}^{2} & = & -\left[\left(1-\frac{2{\mathcal{M}}(u)}{r_{v}}\right)\dot{u}^{2}+2\dot{u}\dot{r}_{v}\right]dt^{2}+r_{v}^{2}d\Omega^{2},\nonumber \\
\end{eqnarray}
where $\dot{}\equiv d/dt$. Matching the induced metrics across $\Sigma$
we get
\begin{equation}
r_{v}(t)=r_{\Sigma}a(t),~~~\left(1-\frac{2{\mathcal{M}}(u)}{r_{v}}\right)\dot{u}^{2}+2\dot{u}\dot{r}_{v}=1.\label{main}
\end{equation}
Matching the extrinsic curvature components calculated from the interior
and exterior geometries and after a straightforward but lengthy calculation,
we get at the boundary\cite{Trends-Bo}
\begin{equation}
2{\mathcal{M}}|_{\Sigma}=r_{\Sigma}^{3}a\dot{a}^{2}=2M|_{\Sigma}.\label{EXMASS}
\end{equation}
Following \cite{Chinese}, let us cast the exterior line element into
dual-null form as
\begin{equation}
ds_{{\rm out}}^{2}=-2d\xi^{+}d\xi^{-}+r_{v}^{2}d\Omega^{2},\label{EXDU}
\end{equation}
with the dual-null one-forms given by
\begin{equation}
d\xi^{+}=\frac{1}{2}du,~~~~~~d\xi^{-}=\left(1-\frac{2{\mathcal{M}}}{r_{v}}\right)du-2dr_{v}.\label{null1}
\end{equation}
The corresponding radial null expansions are given by
\begin{equation}
\theta_{+}=\frac{2}{r_{v}}\left(1-\frac{2{\mathcal{M}}}{r_{v}}\right),~~~\theta_{-}=-\frac{1}{r_{v}}.\label{RNDE}
\end{equation}
The dynamical horizon in the generalized Vaidya spacetime is located
at $\theta_{+}=0$ or simply $2{\mathcal{M}}=r_{v}$, which lies on
the boundary surface if $2M=R$. Then, from \ref{EXMASS} we readily
get
\begin{equation}
|\dot{a}|=\frac{1}{r_{\Sigma}},\label{APH}
\end{equation}
which implies that once the collapse velocity reaches the value that
satisfies the above equation, the dynamical horizon intersects the
boundary of the star. Figure \ref{DPS4ho} shows the absolute value
of the collapse velocity versus the scale factor. The horizontal arrows
label different values of $|\dot{a}|$ for different boundary radii,
as Eq. (\ref{APH}) dictates. There are two thresholds for the
horizon formation, one in the collapse phase, which corresponds to
$|\dot{a}_{{\rm 1max}}|=1/r_{1\Sigma}$, and the other one in the 
expanding phase which corresponds to $|\dot{a}_{{\rm 2max}}|=1/r_{2\Sigma}$.
Thus, for $\ell<0$ (solid curve), the following considerations can
be remarkable:
\begin{itemize}
\item The regularity condition demands that there must be no trapping of
light at the initial epoch from which the collapse scenario begins.
Thus there exists a maximum radius, namely, $r_{{\rm rg\Sigma}}$,
so that if $r_{{\rm b}}=r_{{\rm rg\Sigma}}$ the regularity condition
breaks down. Then, we could deduce that if the boundary surface is
taken so that $r_{1\Sigma}<r_{{\rm b}}<r_{{\rm rg\Sigma}}$ or equivalently
$|\dot{a}_{{\rm rg}}|<|\dot{a}|<|\dot{a}_{{\rm 1max}}|$, three horizons
may appear (see the dashed arrow labeled as D, in Fig.~\ref{DPS4ho});
in the accelerated contracting regime, (from the initial configuration
until point A or the first inflection point, see Fig. \ref{DPS4a})
as $|\dot{a}|$ increases, the first horizon forms to intersect the
boundary until the time at which the decelerated contracting regime
begins. After this time, $|\dot{a}|$ starts decreasing until getting
vanished at the bounce, i.e., from point A to B. During this time
interval where a decelerated contracting regime governs the collapse
procedure, the horizon condition (\ref{APH}) is satisfied for the
second time at an inner horizon. Contrary to the outer horizon, this
one is situated in a modified regime where the weak energy condition
(WEC) is effectively violated (due to the appearance of negative pressure)
and $\ddot{a}>0$. As the collapse is replaced by a bounce and an
accelerated expanding regime gets started, i.e., from points B to
C, the condition (\ref{APH}) is fulfilled for the third time and
a dynamical horizon intersects the matching surface.
\item If $r_{2\Sigma}<r_{{\rm b}}<r_{1\Sigma}$ or equivalently $|\dot{a}_{{\rm 1max}}|<|\dot{a}|<|\dot{a}_{{\rm 2max}}|$,
then no horizon may form in the exterior zone throughout the collapse
regime (see the dashed arrow labeled as E, in figure~\ref{DPS4ho}).
However, two dynamical horizons may still occur to meet the boundary;
one in the accelerated expanding phase ($t_{{\rm cr}}<t<t_{{\rm 2inf}}$)
and the other one in the decelerated expanding phase ($t>t_{{\rm 2inf}}$).
Therefore, the collapse procedure that is replaced by a bouncing scenario
may be covered by these horizons.
\item Finally, if $r_{{\rm b}}<r_{2\Sigma}$ or equivalently $|\dot{a}|>|\dot{a}_{{\rm 2max}}|$,
no horizon may occur in the exterior Vaidya region and the bounce
is uncovered (see the dashed arrow labeled as F, in Fig.~\ref{DPS4ho}).
\end{itemize}
In contrast to the case $\ell<0$ for which $|\dot{a}|$ is bounded
during the evolution of the setting, it grows boundlessly for $\ell>0$
and $\ell=0$ so that there cannot be found any threshold for the
collapse velocity or any minimum radius for the boundary in order
to avoid the formation of horizons. Thus as we approach the singularity,
a dynamical horizon will always form to cover the singularity. In
order to see whether the outward flux of energy can be visible to
faraway observers, we assume that the energy flux as measured locally
by an observer with a four-velocity vector $\zeta^{\mu}$ is given
by \cite{LINSCHM}
\begin{equation}
\sigma\equiv T_{\mu\nu}\zeta^{\mu}\zeta^{\nu}.\label{flux}
\end{equation}
We consider only radially moving observers and define the radial velocity
for such an observer as
\begin{equation}
v\equiv\zeta^{r_{v}}=\frac{dr_{v}}{dt}.\label{radial}
\end{equation}
This follows then from $\zeta_{\mu}\zeta^{\mu}=-1$ and $\zeta^{\theta}=\zeta^{\phi}=0$
\begin{equation}
\frac{du}{dt}=\zeta^{u}=\frac{\eta-v}{\left(1-\frac{2{\mathcal{M}}(u)}{r_{v}}\right)}=\frac{1}{\eta+v},\label{radial1}
\end{equation}
where
\begin{equation}
\eta=\left(1+v^{2}-\frac{2{\mathcal{M}}(u)}{r_{v}}\right)^{-1}.\label{radial2}
\end{equation}
By calculating the nonvanishing component of Ricci tensor, i.e., $\left(R_{uu}=-\frac{2}{r_{v}^{2}}\frac{d{\mathcal{M}}}{du}\right)$
and using Eq. (\ref{flux}), we get the following expression
for the energy flux $\sigma$, as
\begin{equation}
\sigma=-\frac{1}{(\eta+v)^{2}}\left(\frac{1}{4\pi r_{v}^{2}}\frac{d{\mathcal{M}}(u)}{du}\right).\label{sigma}
\end{equation}
The total luminosity for an observer with speed $v$ and the radius
$r_{v}$ is given by \cite{LINSCHM}
\begin{equation}
L(u)=4\pi r_{v}^{2}\sigma.\label{luminosity1}
\end{equation}
Substituting Eq. (\ref{sigma}) into (\ref{luminosity1}) we
get
\begin{equation}
L(u)=-\frac{1}{(\eta+v)^{2}}\frac{d{\mathcal{M}}(u)}{du}.\label{luminosity2}
\end{equation}
Then, using Eq. (\ref{radial1}) in Eq. (\ref{luminosity2}),
we can rewrite the luminosity, in terms of the interior mass function
as
\begin{equation}
L(u)=-\frac{\dot{M}}{(\eta+v)}.\label{luminosity3}
\end{equation}
For an observer being at rest ($v=0$) at infinity ($r_{v}\rightarrow~\infty$),
the total luminosity of the radiation can be obtained by taking the
limit of (\ref{luminosity3}) as
\begin{equation}
L_{\infty}(u)=-\dot{M}.\label{luminosity4}
\end{equation}
As we have described herein this paper, the negative pressure coming
from phase-space deformation effects decelerates the collapse procedure
until the bouncing time. Therefore, when the collapse enters the \textit{slow
reacting} regime, i.e., $t_{{\rm 1inf}}<t<t_{{\rm cr}}$, (see the
lower plot in Fig. \ref{DPS4a}) the horizon is expected to shrink
due to the modifications coming from phase-space deformation in the
interior spacetime, which led to the violation of WEC (this allows
the bounce to happen \cite{Trends-Bo}). Since $\dot{M}<0$ in this
regime we may conclude that $L_{\infty}(u)\!\!\mid_{t_{{\rm 1inf}}<t<t_{{\rm cr}}}>0$;
thus the radiation emanating from the bounce process may be possible
to be detected by external observers. However, since $|\dot{a}|$
is bounded (for $\ell<0$), then, by suitable choice of the boundary
surface $r=r_{\Sigma}$, the horizon formation is avoided. In such a situation,
we have a regular matter configuration that initially collapses, reaches
high densities and then disperses without the horizon formation.

\begin{figure}
\centering\includegraphics[width=3in]{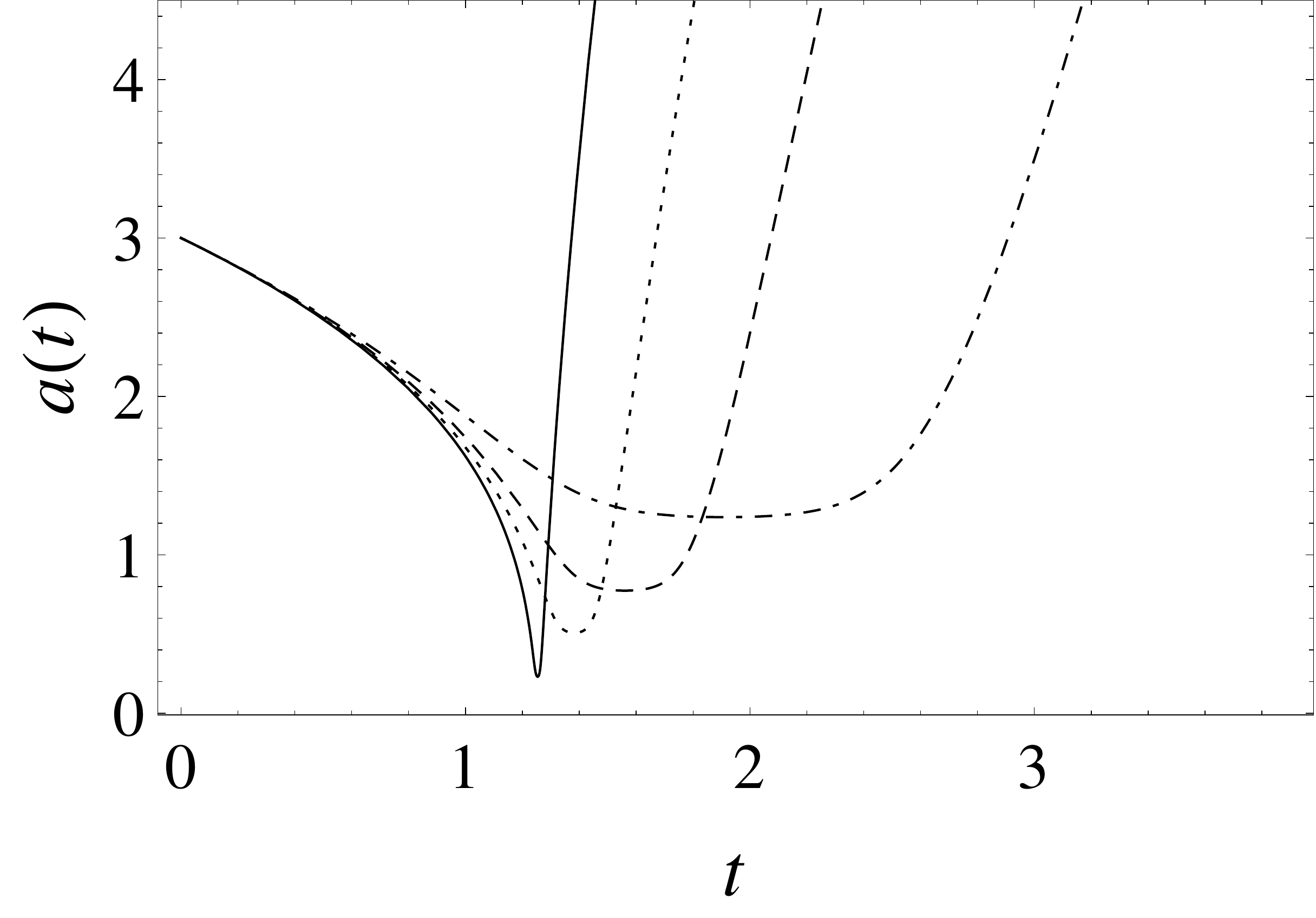}\hspace{4mm} \caption{{\footnotesize{{The time behavior of the scale factor for different values
of deformation parameter, $\ell=-0.07385$ (solid curve), $\ell=-0.1266$
(dotted curve), $\ell=-0.19412$ (dashed curve), $\ell=-0.38402$
(dotted-dashed curve) for $\beta=-3.2$ and $\alpha=1.1$. We have
taken the initial values $\phi(t_{i})=1.98$, $\dot{\phi}(t_{i})=0.711$,
$a(t_{i})=3$, $\dot{a}(t_{i})=-0.868$ and $\rho_{i}=0.2511$.}}}}

\label{DPS4d}
\end{figure}

In Fig.~\ref{DPS4d} we have plotted the scale factor for different
values of deformation parameter. It is seen that, as the absolute
value of $\ell$ increases, the bouncing stage increases. This may
be seen from the pressure originating from the deformation effects
(cf. the upper plot of Fig. \ref{DPS4e}). The larger the value
of $p_{d}$, the longer the time scale of the bounce. Let us briefly
mention that although (\ref{field.eq1}) remains unchanged under deformation,
the time behavior of the kinetic energy of the scalar field in this
equation is different.

\begin{figure}
\centering\includegraphics[width=2.5in]{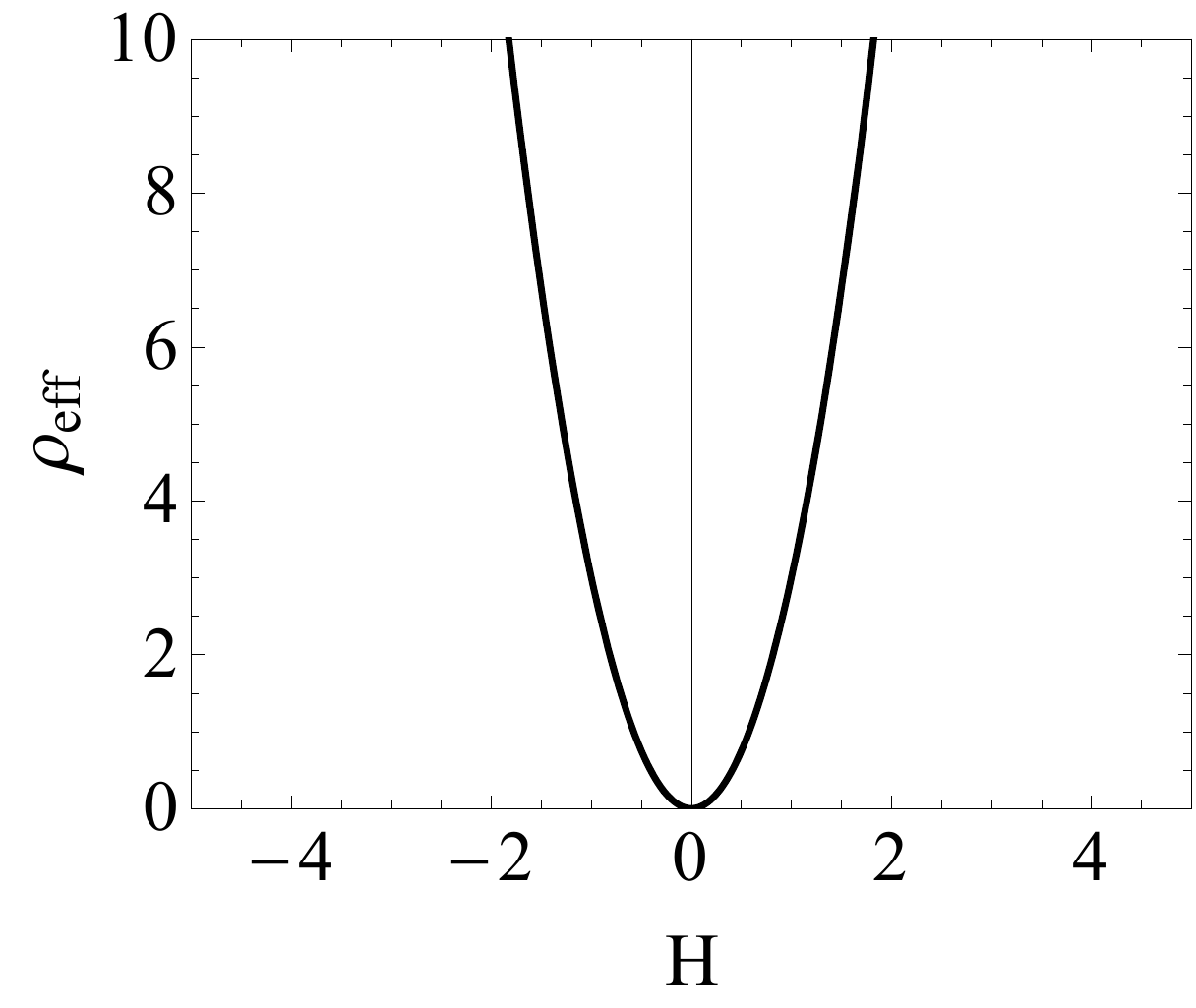} \caption{{\footnotesize{{The phase-space of the collapse rate and effective
energy density for $\ell=-0.211$, $\beta=-3.2$ and $\alpha=1.1$.
We have taken the initial values $\phi(t_{i})=1.98$, $\dot{\phi}(t_{i})=0.711$,
$a(t_{i})=3$, $\dot{a}(t_{i})=-0.868$ and $\rho_{i}=0.2511$.}}}}

\label{rhoh}
\end{figure}

Fig. \ref{DPSphi-phidot}~ shows our numerical simulation for the
scalar field and its kinetic energy. The full curve shows that the
scalar field increases monotonically with a soft slope at the bouncing.
The kinetic energy decreases after a local maximum as the collapse
reaches the bouncing time, while regarding (\ref{HPotential CC}),
the potential energy decreases too and cancels the kinetic energy.
Therefore, the collapse rate vanishes and changes from a collapsing
phase to a bouncing phase \cite{GJM12}. Such a transition can be
better seen in the phase portrait of the collapse rate and effective
energy density. As we see from Eq. (\ref{primed-dps-eq1}) {[}or
from (\ref{dps-eq1}){]}, there are two branches i.e., collapsing
and expanding phases for negative and positive signs of the collapse
rate, respectively. Thus, as the star initially begins to contract,
the collapse rate approaches zero (see the left half-plane of Fig.
\ref{rhoh}), where the bounce appears at a finite scale factor and
then starts to expand at later times, see right half-plane of Fig.
\ref{rhoh}. The dotted-dashed curve in Fig.~\ref{DPSphi-phidot}~~presents
the situation in the absence of phase-space deformation, with the
scalar field increasing boundlessly and the kinetic energy diverging
to infinity. The dashed curve shows the behavior of these quantities
for positive value of $\ell$ whereas we see the kinetic energy grows
more rapidly than the case $\ell=0$.

\begin{figure}
\centering\includegraphics[width=3in]{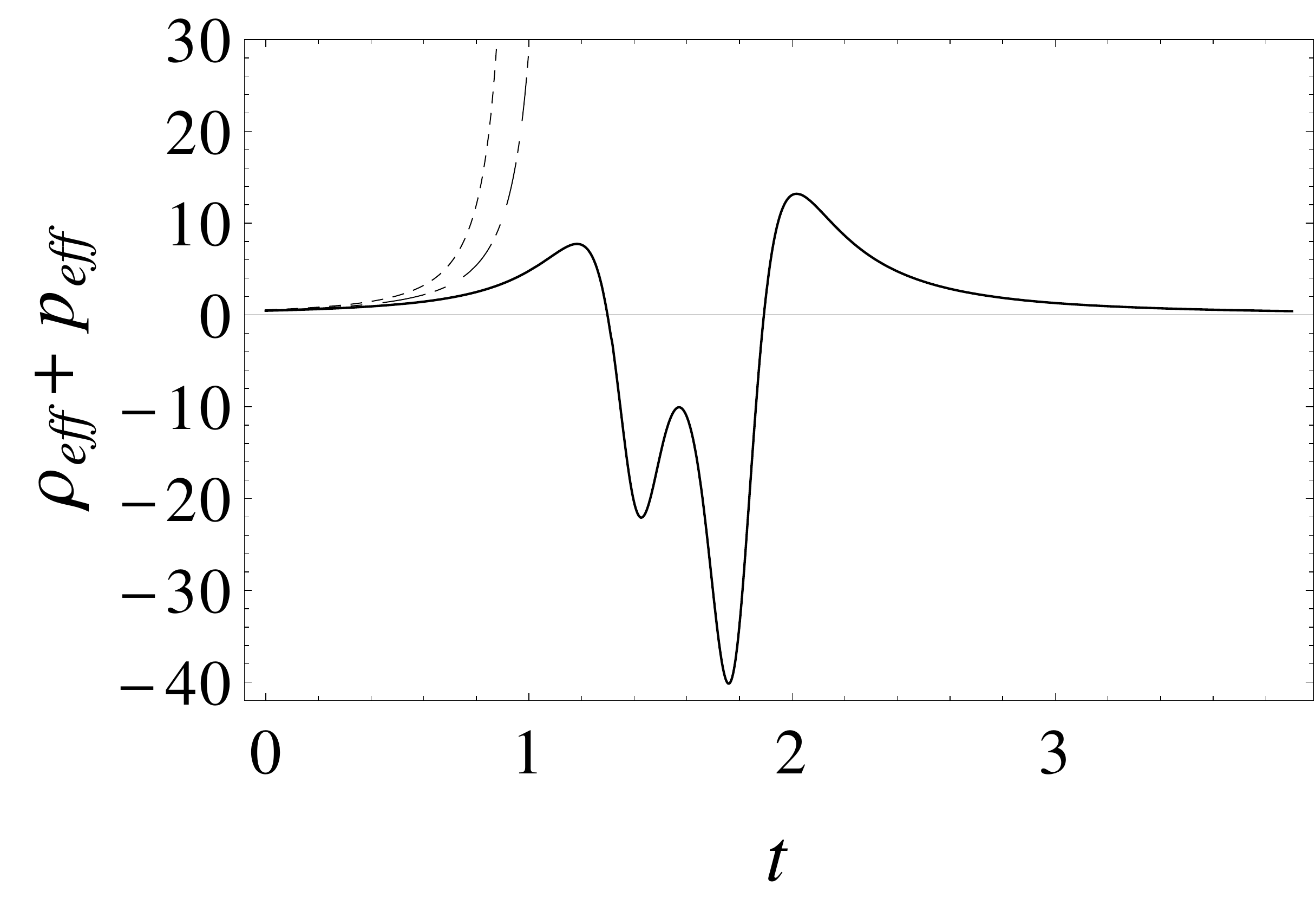} \caption{{\footnotesize{{The behavior of weak energy condition for different
values of deformation parameter, $\ell=-0.211$ (solid curve), $\ell=0.211$
(dashed curve), $\ell=0$ (dotted-dashed curve) for $\beta=-3.2$,
and $\alpha=1.1$. We have taken the initial values $\phi(t_{i})=1.98$,
$\dot{\phi}(t_{i})=0.711$, $a(t_{i})=3$, $\dot{a}(t_{i})=-0.868$
and $\rho_{i}=0.2511$.}}}}

\label{WEC}
\end{figure}

For the collapsing matter clouds, it is usually required that the
WEC be respected by the collapse configuration. In the collapse setting
presented herein, WEC is given by $\rho_{{\rm eff}}\geq0$ and $\rho_{{\rm eff}}+p_{{\rm eff}}\geq0$.
The first inequality holds for both undeformed and deformed configurations
and the second one also remains valid for undeformed case. However,
as Fig. \ref{WEC} shows the second inequality holds only in the
weak field regime, i.e., close to the initial configuration of the
collapse setting, since the pressure appearing as phase-space deformation
effects becomes dominant over the effective energy density at later
stages. Such a crucial feature leads to the violation of WEC, which
can be seen in several collapse settings where the effects of quantum
gravity become prominent in strong field regimes \cite{DMQG}. However
the herein setting guarantees the positivity of the effective energy
density.

\section{Conclusions}

\label{Concl}

In this paper, concerning a scenario of gravitational collapse, we
probed singularity formation (or the possibility of its removal) in
the presence of a phase-space deformation within the canonical momenta
sector. To be more precise, our matter content was described by the
Lagrangian density of a scalar field minimally coupled to the spacetime
curvature. The interior spacetime as taken as that of flat Friedmann-Lema\^{\i}tre-Robertson-Walker
metric \cite{JG04}, \cite{JGS-LQ}, \cite{BOJOetal}. Thereby, by
employing an Hamiltonian formalism, we explored the consequences of
the dynamical deformation~(\ref{deformed}) in the phase-space.

The choice of such a type of deformation, whose particular form can
be further discussed by means of a dimensional analysis, was motivated
\cite{RFK11}. Additional arguments for it can be found also in~\cite{RFK11}
, namely with respect to the noncommutativity between the canonical
momenta%
\footnote{By taking the standard Brans-Dicke Lagrangian in vacuum and a spatially
flat Friedmann-Robertson-Walker metric, then applying a dynamical deformation in the phase
space, the big bang singularity is removed and also the horizon problem
is analyzed \cite{RFK11}.%
}.

More concretely, the phase-space deformation emerges in the equations
of motion by means of specific new terms, characterized by a parameter
$\ell$. This could be taken either as positive or negative ($\ell=0$
representing the no deformation setting). The case $\ell>0$ leads
us to an additional positive pressure effect, speeding up the collapse
toward the singularity. Whereas, in the $\ell<0$ case, a negative
pressure is present, inducing the collapse (that would have been toward
a singularity) to be replaced by a nonsingular bounce.

It may be appropriate to compare the deformed equations with the corresponding
ones when $\ell=0$. In the usual collapsing regime, as is seen from
Eq.~(\ref{field.eq3}), the $\dot{\phi}$ term acts as antifriction
throughout the collapse. From Eq. (\ref{primed-dps-eq3}), we
may intuitively consider that a negative value of a quantity associated
to phase-space deformation parameter, would balance the antifriction
term%
\footnote{The same \textit{antifrictional} behavior can be seen when loop quantum
effects are taken into account in the collapse process of a scalar
field \cite{BOJOetal}.%
}. In fact, it was precisely an additional negative pressure ($p_{d}$),
induced from the phase-space deformation, that changed the collapsing
picture: in the undeformed regime, for $\beta<-1$, trapped surfaces
do form as Eq. (\ref{MSR}) shows and thus the resulting singularity
is covered, while in the deformed one (for $\ell<0$) trapped surfaces
may be avoided till the time at which the bounce occurs, see the middle
plot of Fig. \ref{DPS4b} . Since trapped surfaces are failed to
form, there may exist an outward flux of energy, due to which the
effective mass reduces, see the $\dot{M}<0$ period in the lower plot
of Fig.~\ref{DPS4c} . At later times, when the star begins an
expanding phase, it absorbs the energy that has been escaping the
collapsing phase. Thus, gravity becomes repulsive due to the presence
of $p_{d}$. This provides the bouncing behavior and hence a singularity
avoidance, depending thus on the deformation parameter. It is worth
noting that the time scale of the bounce depends on the absolute magnitude
of the deformation parameter.

\begin{figure}
\centering\includegraphics[width=3in]{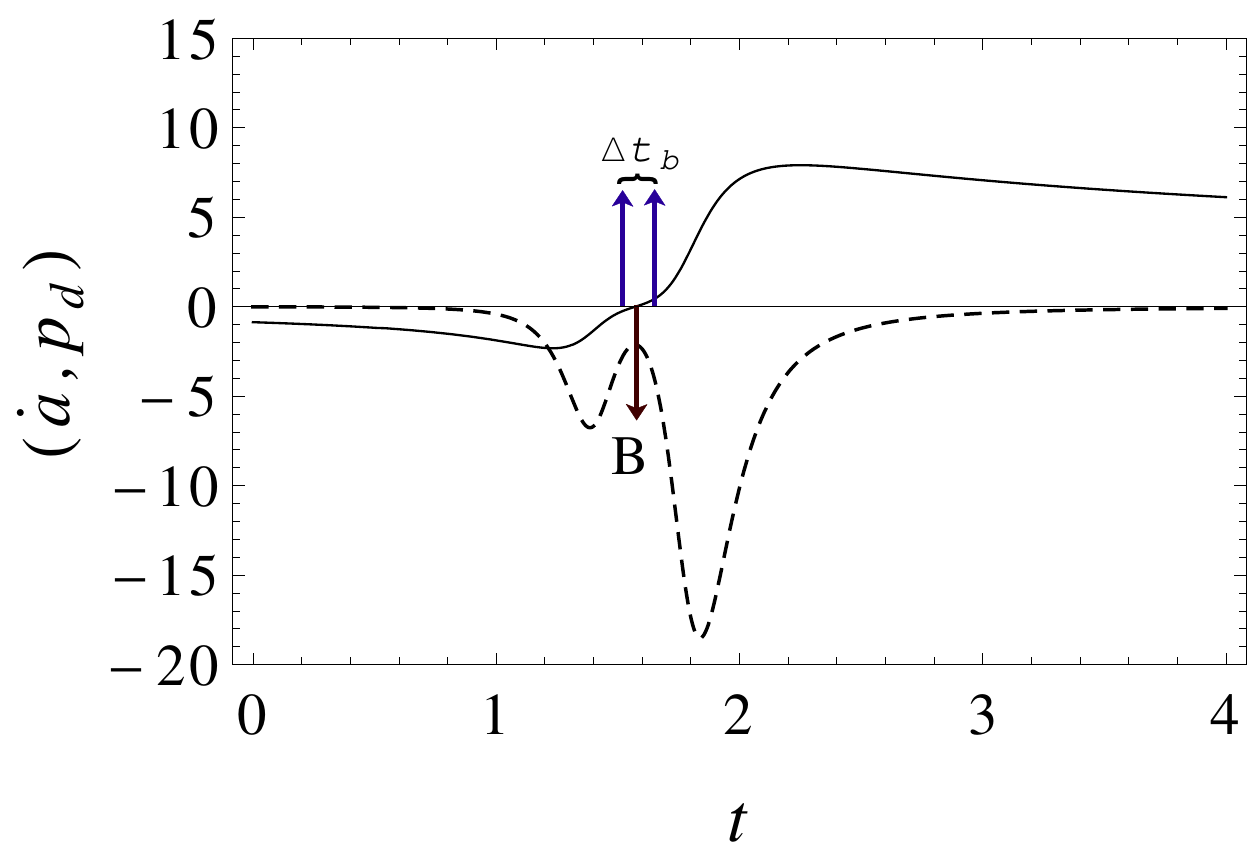} \includegraphics[width=3in]{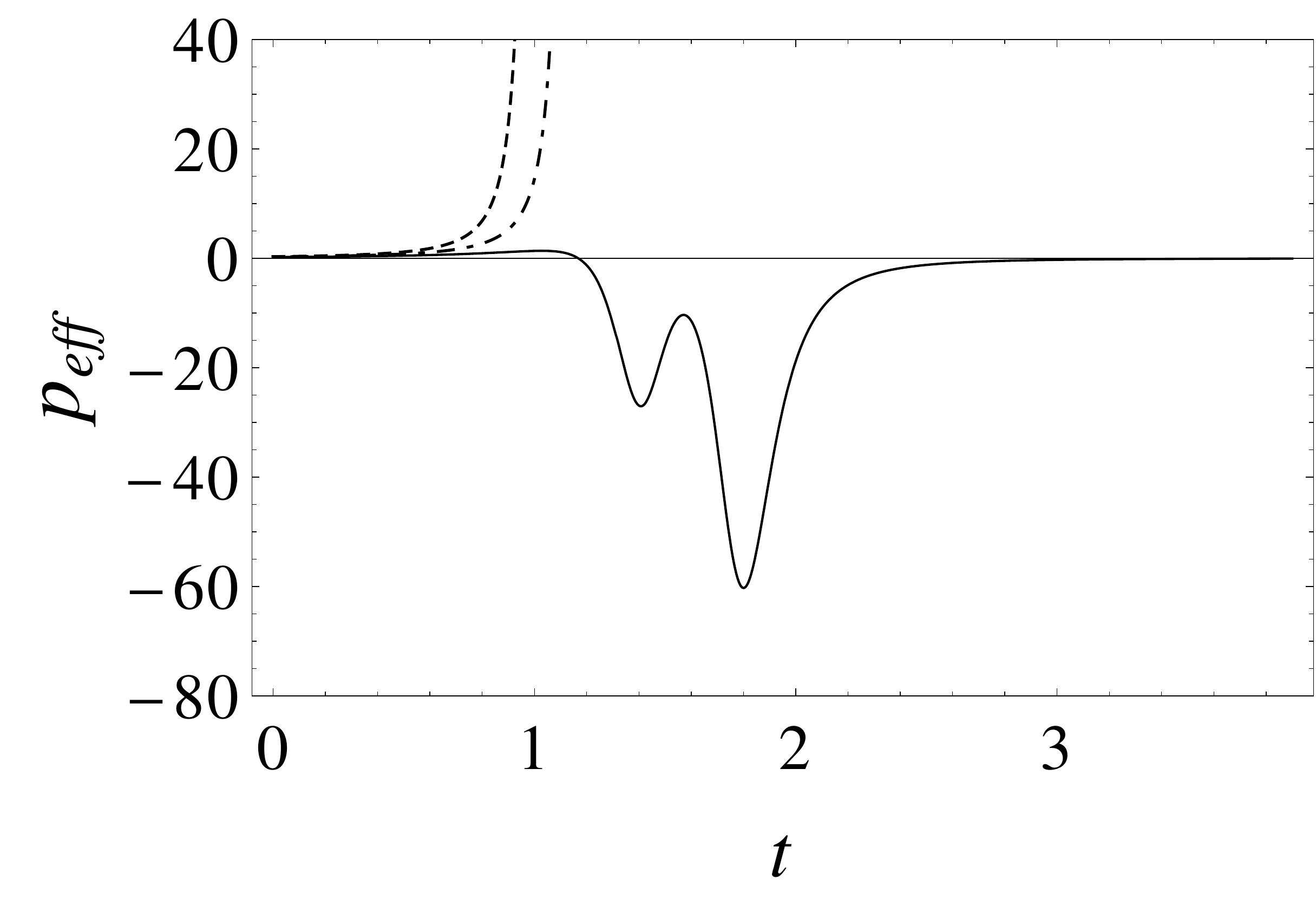}
\includegraphics[width=3in]{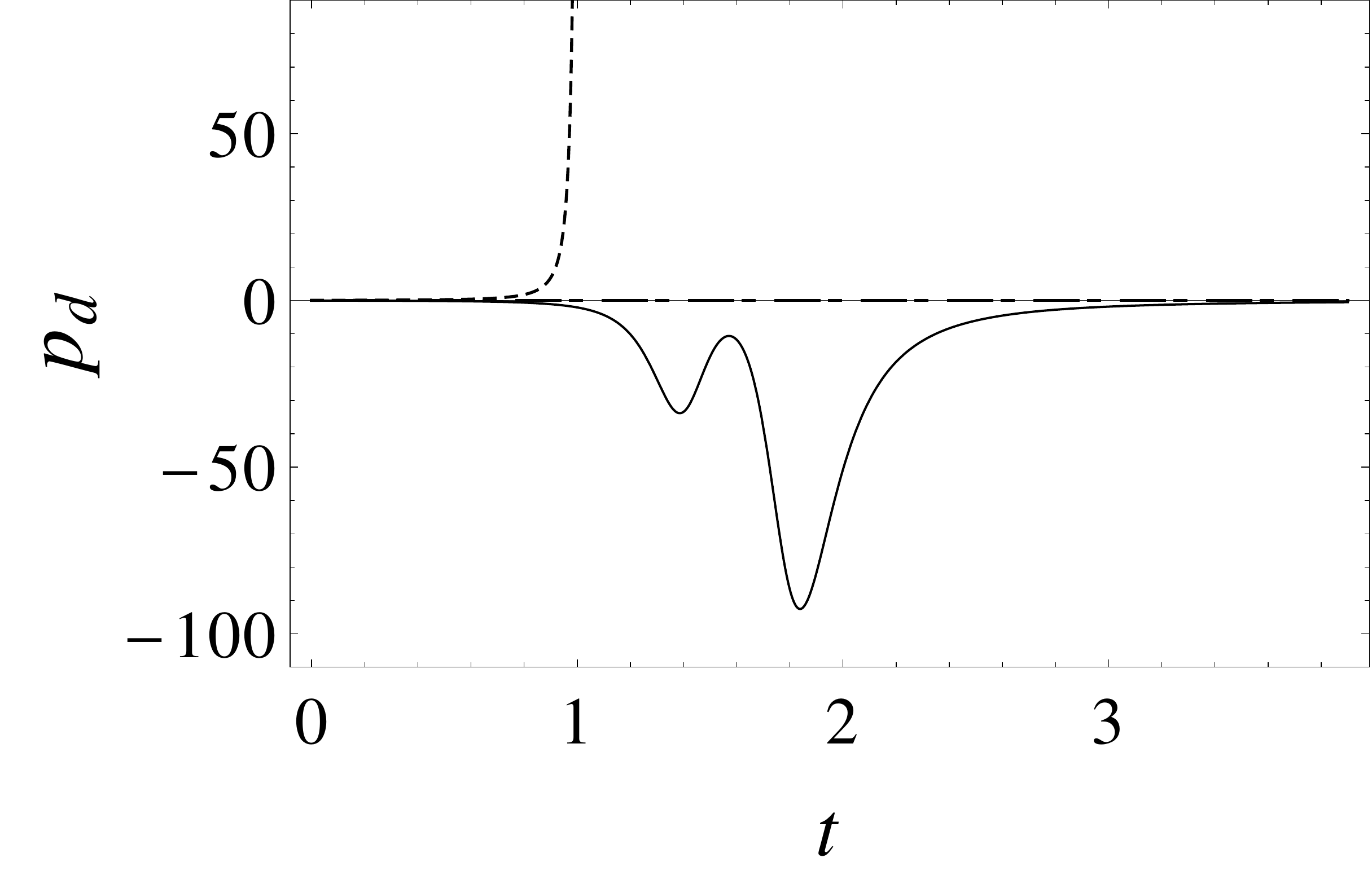} \caption{{\footnotesize{{Upper panel: The time behavior of the induced pressure
originated from the phase-space deformation and the collapse velocity
for $\ell=-0.211$. Middle panel: The time behavior of effective pressure,
for different values of deformation parameter, $\ell=-0.211$ (solid
curve), $\ell=0.211$ (dashed curve), $\ell=0$ (dotted-dashed curve)
for $\beta=-3.2$ and $\alpha=1.1$. Lower panel: The time behavior
of $p_{d}$ for $\ell=-0.211$ (solid curve), $\ell=0.211$ (dashed
curve), $\ell=0$ (dotted-dashed curve), and $\beta=-3.2$. We have
taken the initial values $\phi(t_{i})=1.98$, $\dot{\phi}(t_{i})=0.711$,
$a(t_{i})=3$, $\dot{a}(t_{i})=-0.868$ and $\rho_{i}=0.2511$. }}}}

\label{DPS4e}
\end{figure}

In summary, in the early stages of the collapse, as time advances,
the velocity of the collapse becomes more and more negative; meanwhile
the pressure $p_{d}$ that emerges from phase-space deformation effects
comes into play, to prevent the collapsing phase to proceed. This
pressure starts from a small negative value and progresses gradually
to more negative values, thus ceasing the growth of the collapse velocity,
up to a maximum negative value (see point A in Fig. \ref{DPS4a}).
From then, the collapse continues but in a decelerating phase so that
$p_{d}$ reaches its local maximum in negative direction
(see the upper plot in Fig. \ref{DPS4e}). After that, $p_{d}$
proceeds, competing with gravitational attraction, until the time at
which the collapse velocity becomes zero, where, during $\Delta t_{{\rm b}}$,
the pressure $p_{d}$ stays for a while in its local minimum. Then,
the collapse smoothly transforms to an accelerated expansion owing
to this negative pressure, and this situation continues until that
$p_{d}$ achieves its absolute maximum (negative value), where $\dot{a}$
and the effective energy density reach their maximum value. Finally,
as the procedure enters a weak field regime, the deformation effects
start to ease, so that the velocity of expansion and the effective
energy density converge asymptotically. However, for positive values
of the phase-space deformation parameter, it makes the last term in
Eq. (\ref{primed-dps-eq3}) behave as an antifriction term
and prompts the collapse scenario to reach the singularity faster
than the case in which the deformation effects are absent. The middle
plot of Fig. \ref{DPS4e}~~further shows that the effective pressure
for $\ell<0$ begins from positive values, then turning to negative
ones as the bounce occurs. For $\ell>0$ the effective pressure remains
always positive and diverges. Indeed, the associated positive pressure
adds to gravity and strengthens its attractiveness property, see dashed
curve in the lower plot of Fig. \ref{DPS4e}. Let us further mention
that, from Fig. (\ref{DPS4a}), if we set the interval $\Delta t_{s}=t_{s}-t_{i}$
as the time that the collapse takes to hit the singularity, then the
following inequality holds
\begin{equation}
\Delta t_{\ell^{+}}<\Delta t_{s}<\Delta t_{\ell^{-}},\label{TI}
\end{equation}
where $\Delta t_{\ell^{+}}=\bar{t}_{s}-t_{i}$ is the time that the
collapse scenario takes to reach the singularity at $t=\bar{t}_{s}$
(for $\ell>0$), and $\Delta t_{\ell^{-}}=t_{{\rm cr}}-t_{i}$ is
the time that is taken up until the collapse transforms to a bounce.
The fact is that for $\ell<0$, the collapse slows down due to the
appearance of the negative pressure $p_{d}$, which prompts the collapse
to turn into a bounce at the time $t_{{\rm cr}}>t_{s}$. Conversely,
for $\ell>0$, the corresponding positive pressure causes the collapse
to reach the singularity at an earlier time than the case $\ell=0$;
hence $\bar{t}_{s}<t_{s}$.

Finally, we would like to point out that besides the setting presented
here, there are various works in the literature on other bouncing
scenarios. Among them we quote $f(R)$ theories in Palatini formalism
\cite{FRPal}, generalized teleparallel gravity theories \cite{TELL}
and in the presence of interacting spinning particles in the framework
of Einstein-Cartan theory \cite{ECSPIN}. The occurrence of bounces
have been reported in spatially flat isotropic models in loop quantum
cosmology for a massless scalar field \cite{LQCSFH}, for different
matter models \cite{LQCMA}, and in the presence of anisotropy \cite{LQCANISO}
(see also \cite{LQCUP} and references therein). In addition, models
based on loop quantum gravity suggest that the singularity (that forms
in the classical framework of gravitational collapse), can be regularized
when the collapse scenario enters the Planckian regimes; semiclassical
effects into the gravitational collapse of a homogeneous scalar field
replace the singularity by a nonsingular bounce \cite{BOJOetal,YJA}.
The same approach for a closed universe filled with a massive scalar
field, which classically collapses to a singularity has been investigated
in \cite{BIGC} and it was shown that loop quantum effects in high
curvature regimes led to a bouncing scenario, irrespective of the
initial conditions. In the end, coordinate noncommutativity may also
result in remarkable cosmological scenarios \cite{Octavio-COS} as
well as its role in curing the problems we face in describing the
final fate of a radiating black hole, such as removing the curvature
singularity being present in the commutative case \cite{Nicolini}.

\section{Acknowledgments}

One of us (S.M.M.R) is grateful for the support of Grant No. SFRH/BPD/82479/2011
from the Portuguese Agency Funda\c{c}\~ao para a Ci\^encia e Tecnologia.
This research work was supported by Grant No. CERN/FP/123618/2011.
We would like to thank Nima Khosravi and Shahram Jalalzadeh for useful
discussions and comments.

\appendix

\section{Another (second) approach for deriving the equations of motion of
sec. (III)}

\label{App.A}

Here, we would like to apply another approach, which has been employed
in the former investigations (see e.g. Refs.~\cite{Other},\cite{KS08}),
to derive the equations of motion associated to the deformed space
of Sec.~\ref{DPS}. Let us start by introducing the following
variables
\begin{eqnarray}
\begin{cases}
P'_{\phi'}\,=\, P_{\phi}-\ell a\phi^{3}\\
\phi'(t)=\,\phi(t),\; N'(t)=N(t)\\
a'(t)\,=\, a(t),\; P'_{a'}=P_{a}.
\end{cases}\label{trick}
\end{eqnarray}
We can easily show that the above variables satisfy the relation~(\ref{deformed})
if the unprimed variables satisfy the standard Poisson brackets. By
employing the above transformations, the Hamiltonian~(\ref{Ham.prime})
changes to
\begin{equation}
{\mathcal{H}}_{0}^{{\rm nc}}={\mathcal{H}}_{0}-N\ell a^{-2}\phi^{3}P_{\phi}+\frac{1}{2}N\ell^{2}a^{-1}\phi^{6},\label{NewDirac.H}
\end{equation}
where $H_{0}$ is given by (\ref{H0}). In fact, by employing the
transformation~(\ref{trick}), the Hamiltonian ${\mathcal{H'}}_{0}$
(as a function of the primed variables) has been replaced by ${\mathcal{H}}_{0}^{{\rm nc}}$,
as a function of the unprimed variables. Finally, we write the Dirac
Hamiltonian for the deformed scenario as
\begin{equation}
{\mathcal{H}}^{{\rm nc}}={\mathcal{H}}_{0}^{{\rm nc}}+\lambda P_{N}.\label{NewDirac.H-a}
\end{equation}

The equations of motion with respect to the above Hamiltonian become
\begin{eqnarray}
\dot{a}\!\! & = & \!\!\!\{a,{\mathcal{H}}^{{\rm nc}}\}=-\frac{1}{6}Na^{-1}P_{a},\label{mot.eq1}\\
\dot{P}_{a}\!\! & = & \!\!\!\{P_{a},{\mathcal{H}}^{{\rm nc}}\}=-\frac{1}{12}Na^{-2}P_{a}^{2}+\frac{3}{2}Na^{-4}P_{\phi}^{2}\nonumber \\
\!\! & - & \!\!\!2N\ell a^{-3}\phi^{3}P_{\phi}+\frac{1}{2}N\ell^{2}a^{-2}\phi^{6}-3Na^{2}V(\phi),\label{mot.eq2}\\
\dot{\phi}\!\! & = & \!\!\!\{\phi,{\mathcal{H}}^{{\rm nc}}\}=Na^{-3}P_{\phi}-N\ell a^{-2}\phi^{3},\label{mot.eq3}\\
\dot{P}_{\phi}\!\! & = & \!\!\!\{P_{\phi},{\mathcal{H}}^{{\rm nc}}\}=3N\ell a^{-2}\phi^{2}P_{\phi}-3N\ell^{2}a^{-1}\phi^{5}\nonumber \\
 & - & Na^{3}\frac{dV(\phi)}{d\phi},\label{mot.eq4}\\
\dot{N}\!\! & = & \!\!\!\{N,{\mathcal{H}}^{{\rm nc}}\}=\lambda\,,\label{mot.eq5}\\
\dot{P}_{N}\!\! & = & \!\!\!\{P_{N},{\mathcal{H}}^{{\rm nc}}\}=\frac{1}{12}a^{-1}P_{a}^{2}-\frac{1}{2}a^{-3}P_{\phi}^{2}\nonumber \\
\!\! & + & \!\!\!\ell a^{-2}\phi^{3}P_{\phi}-\frac{1}{2}\ell^{2}a^{-1}\phi^{6}-a^{3}V(\phi),\label{mot.eq6}
\end{eqnarray}
where, to derive the above equations, we have used the ordinary (standard)
Poisson brackets. However, instead of the standard Hamiltonian~(\ref{Dirac.H}),
the noncommutative/deformed Hamiltonian~(\ref{NewDirac.H-a}) has
been employed. Therefore, we should mention that the unprimed variables
in this section are perfectly different from their corresponding in
Sec. \ref{Standard}, and they were denoted by the present shapes
(i.e., in unprimed forms) only for simplicity.

In the comoving gauge, it is straightforward to show that the equations
of motion are given by
\begin{eqnarray}
\left(\frac{\dot{a}}{a}\right)^{2}=\frac{1}{3}\left[\frac{1}{2}\dot{\phi}^{2}+V(\phi)\right]=\frac{1}{3}\rho_{{\rm eff}},\label{dps-eq1}
\end{eqnarray}
\begin{eqnarray}
2\frac{\ddot{a}}{a}+\left(\frac{\dot{a}}{a}\right)^{2}\!\!\! & = & \!\!\!-\left[\frac{1}{2}\dot{\phi}^{2}-V(\phi)\right]-\frac{1}{3}\ell a^{-2}\phi^{3}\dot{\phi}\nonumber \\
\!\!\! & \equiv & \!\!\!-(p+p_{{\rm d}})\equiv-p_{{\rm eff}},\label{dps-eq2}
\end{eqnarray}
\begin{equation}
\ddot{\phi}+3\left(\frac{\dot{a}}{a}\right)\dot{\phi}+\frac{dV(\phi)}{d\phi}+\ell\dot{a}\left(\frac{\phi}{a}\right)^{3}=0,\label{dps-eq3}
\end{equation}
where $p_{{\rm d}}\equiv1/3\ell a^{-2}\phi^{3}\dot{\phi}$.

In order to monitor the role of transformations~(\ref{trick}) more
vividly in this paper, we should mention a few comments regarding the
primed and unprimed variables. The behavior of the unprimed variables
in this section, and also in Sec.~\ref{DPS}, are not the same
as the behavior of the corresponding ones in Sec.~\ref{Standard}.
As mentioned several times, they reduce to their standard counterparts
by letting $\ell=0$. In fact, in this paper, we have obtained the
equations of motion by means of two different approaches and realized
that these equations are perfectly equivalent. In fact, introducing
relations such as~(\ref{trick}), that can be seen in many investigations
{[}see e.g.~\cite{Other}\cite{KS08}{]}, may just be an appropriate
mathematical transformation, at least in our paper, to derive the
equations of motion with a simpler manner. We should stress that the
only way to recover the standard (commutative) results from the noncommutative
ones is to set the deformation parameter equal to zero.\\
 \\

\section{About the equations of motion in Sec. III}

\label{App.B}

The effects of phase-space deformation employed in this paper reveals
itself as an additional pressure in Eq. (\ref{dps-eq2}) so that
the conservation of the effective energy-momentum tensor \cite{Octavio-COS}
leads to the modified evolution equation for the scalar field. Let
us take the derivative of the right- and left-hand side of Eq.~(\ref{dps-eq1}),
giving
\begin{eqnarray}
\dot{\rho}_{{\rm eff}}=6H\dot{H}.\label{Check-eq1}
\end{eqnarray}
From Eqs.~(\ref{dps-eq1}) and ~(\ref{dps-eq2}), we have
\begin{eqnarray}
\rho_{{\rm eff}}+p_{{\rm eff}}=2H^{2}-2\frac{\ddot{a}}{a}.\label{check-eq2}
\end{eqnarray}
Applying Eqs.~(\ref{Check-eq1}), (\ref{check-eq2}) and $\dot{H}=\ddot{a}/a-H^{2}$,
it provides
\begin{eqnarray}
\dot{\rho}_{{\rm eff}}+3H\left(\rho_{{\rm eff}}+p_{{\rm eff}}\right)=0.\label{Check-eq3}
\end{eqnarray}
We now use the right-hand side and middle expression of Eq.~(\ref{dps-eq1})
to obtain
\begin{eqnarray}
\dot{\rho}_{{\rm eff}}=\dot{\phi}\ddot{\phi}+\dot{\phi}\frac{dV(\phi)}{d\phi}.\label{Check-eq4}
\end{eqnarray}
Again, using the right-hand side and middle expresions of Eqs.~(\ref{dps-eq1})
and ~(\ref{dps-eq2}), gives
\begin{eqnarray}
3H\left(\rho_{{\rm eff}}+p_{{\rm eff}}\right)=3H\dot{\phi}\left(\dot{\phi}+\frac{1}{3}\ell a^{-2}\phi^{3}\right).\label{Check-eq5}
\end{eqnarray}
Adding the left- and right-hand sides of Eqs.~(\ref{Check-eq4}) to
~(\ref{Check-eq5}), it follows
\begin{eqnarray}
 & \dot{\rho}_{{\rm eff}} & +~3H\left(\rho_{{\rm eff}}+p_{{\rm eff}}\right)\\
 & = & \dot{\phi}\left(\ddot{\phi}+\frac{dV(\phi)}{d\phi}+3H\dot{\phi}+\ell\dot{a}a^{-3}\phi^{3}\right).\nonumber
\end{eqnarray}
By assuming $\dot{\phi}\neq0$, we can obtain
\begin{equation}
\ddot{\phi}+3\left(\frac{\dot{a}}{a}\right)\dot{\phi}+\frac{dV(\phi)}{d\phi}+\ell\dot{a}\left(\frac{\phi}{a}\right)^{3}=0.\label{Check-eq6}
\end{equation}
Obviously, all of the explanations of this section are also valid for
the primed variables of Sec.~\ref{DPS}.

\end{document}